\documentclass[sigchi, review]{acmart}

\usepackage{booktabs} 
\usepackage{color}
\usepackage{balance}
\setcopyright{none}

\acmDOI{}

\acmISBN{Author copyright, 2018.}

\acmConference[First completed on 21 September, 2018]{}{in San Jose, CA, USA}

\begin{document}
\title{Mobile Head Tracking for eCommerce and Beyond}

\author{Muratcan Cicek}
\orcid{1234-5678-9012}
\affiliation{%
  \institution{University of California, Santa Cruz}
  \streetaddress{P.O. Box 1212}
  \city{Santa Cruz}
  \state{California}
  \postcode{43017-6221}
}
\email{mcicek@ucsc.edu}

\author{Jinrong Xie}
\affiliation{%
  \institution{eBay Inc.}
  \streetaddress{P.O. Box 1212}
  \city{San Jose}
  \state{California}
  \postcode{43017-6221}
}
\email{jinnxie@ebay.com}

\author{Qiaosong Wang}
\affiliation{%
  \institution{eBay Inc.}
  \streetaddress{1 Th{\o}rv{\"a}ld Circle}
  \city{San Jose}
  \country{California}
}

\email{qiaowang@ebay.com}

\author{Robinson Piramuthu}
\affiliation{%
  \institution{eBay Inc.}
  \city{San Jose}
  \country{California}
}
\email{rpiramuthu@ebay.com}

\renewcommand{\shortauthors}{M. Cicek et al.}

\begin{abstract}
Online shopping is difficult for people with motor impairments. Proprietary software can emulate mouse and keyboard via head tracking. Smartphones are easier to carry indoors and outdoors compared to bulky laptop or desktop devices. However, head tracking solutions are not common for smartphones. To address this, we implement and open source button that is sensitive to head movements tracked from the front camera of iPhone X. This allows developers to integrate in eCommerce applications easily without requiring specialized knowledge. Other applications include gaming and use in hands-free situations such as during cooking, auto-repair. We built a sample online shopping application that allows users to easily browse between items from various categories and take relevant action just by head movements. We present results of user studies on this sample application and also include sensitivity studies based on two independent tests performed at 3 different distances to the screen.
\end{abstract}

%
%



\maketitle

\section{Introduction}
\begin{figure}[t]
\centering
\begin{tabular}{ccc}
    \includegraphics[width=0.25\linewidth]{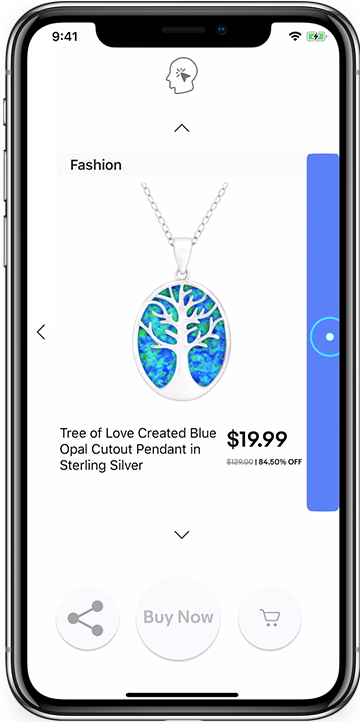} &   \includegraphics[width=0.25\linewidth]{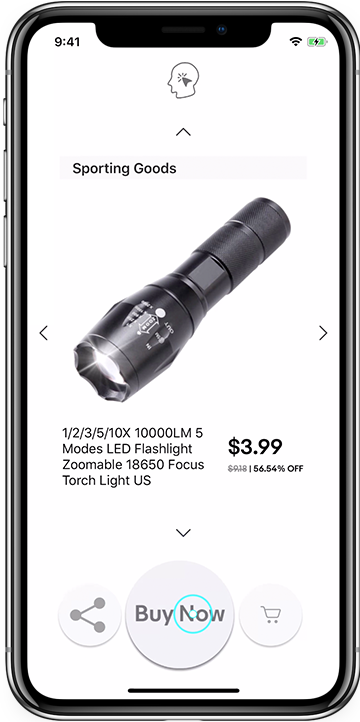} 

\end{tabular}
\caption{Illustration of a sample eCommerce application that we designed and open sourced. Cursor is shown as a cyan circle with dot at its center. Navigation buttons are filled with blue when they are fully activated (left image). Action buttons expand fully for activation (right image).}
\label{fig:Teaser}
\end{figure}

According to Pew Research Center in 2016 \cite{smith2016online} 80\% of Americans shop online. Shopping at stores can be cumbersome as consumers do not prefer to be present at stores physically for several reasons \cite{kpmg2017truth}. In addition, there are also specific barriers \cite{swaine2014exploring} that make in-store shopping harder for people with motor impairments. We believe that online shopping would address this problem since it requires considerably less effort to interact with a digital devices to buy an item than to travel to the nearest store. But even this task is hard to complete for people with motor impairments. They are required to operate a pointer with a mouse-like device or tapping a certain point on the touch screen \cite{findlater2017comparing, montague2014motor, riviere1996effects, wobbrock2014improving}. According to The Centers for Disease Control and Prevention, there are 39.5 million adult Americans with some sort of physical functioning difficulty \cite{disabilities} and they potentially have difficulties during shopping, both in-store and online. 

Proprietary software can emulate mouse and keyboard via head tracking. However, such solution is not common on smartphones. Modern smartphones such as iPhone X are packed with various sensors and abundant computing power. Unlike desktop and laptop computers, they are also much easier to carry around. Therefore, it is natural to make head tracking based interaction available on smartphones. For example, a button, which is typically sensitive to touch, could capture head tracking motion and activate when tracking region falls within its extent. Currently, it is difficult for developers to achieve this since it requires special expertise. Our hope is to make it easy for application developers to use buttons that capture head tracking events. This will make basic activities such as shopping easily accessible to people with motor impairments. Also, this could be extended to other situations demanding hand-free browsing. 

We create such a button and plan to open source the project so that developers could easily integrate to their own projects. Using our software, users could interact with the application only by head movement. This requires significantly less effort compared to using touchscreen. We build our application on top of the state-of-the-art ARKit 2 \cite{arkit} toolkit. Our work shows that the Augmented Reality tools can be used for assistive technology. 

In this paper, we focus on its application as an assistive tool mainly for people with motor impairments. To illustrate the use of such modality, we provide an online shopping experience for everyone through a mobile application. Although our initial motivation was developing an accessible online shopping application, the scope of possible use cases includes other applications such as gaming and hands-free situations such as during cooking, auto-repair. A great variety of mobile applications can be empowered with a touch-free design by employing our ready-to-use interface module.


\section{KEY CONTRIBUTIONS}
The contributions of this work are: 
\begin{itemize}  
  \item An open-source solution for mobile user interfaces that provides hands-free interaction with smartphones. The code is completely open-sourced and will be available soon. It features an accurate mapping from 3D head pose to a virtual pointer, a dwelling function to generate hands-free touch events, a sample practice interface and a demo application for online shopping experience. Anyone should be able to extend our work and/or develop hands-free applications by integrating their design with our UI components.
  
  \item Design of a mobile user interface for online shopping that can benefit everyone including the users with motor impairments. To the best of our knowledge, this research is the first one that highlights the defined problem and proposes a practical solution to developing a working product. 
\end{itemize}
\balance
\section{Related Works}
Our goal is to make basic activities such as shopping to be accessible to everyone via smartphone, including people with motor impairments. We first revisit most of the related work on online shopping and walk through several well-known eCommerce platforms prior to our study. In addition, we will also review interface requirements that are employed in online shopping and existing methods that try to meet these requirements.

\subsection{Assistive Technology}
WebAIM \cite{webaim} lists several \textit{selection} and \textit{pointing} methods as the mainstream assistive technologies. 

\subsubsection{Selection Methods:} Adaptive switches with any kind replace a necessary button pressing for mouse clicking, tapping or for a more specific function like in use of a screen scanning solution. By design, adaptive switches are larger buttons that can be attached anywhere from floor to wheelchair headrests and can be hit by appropriate body parts. Tecla \cite{teclaswitch} lists 7 most common switches as adaptive switch, including several types of buttons, joysticks, Sip-and-Puff tools and blink recognizers. 
The mobile platform, iOS by Apple \cite{switchcontrol} where we develop our application also has built-in features like touch screen tapping as a switch and a front camera-based switch that detects very simple head movements as trigger in addition to the support of external switches. 

\textit{Dwelling function} is a selection mechanism rather than an assistive tool by itself. Most of solutions tend to have dwelling function as a replacement for switches we mentioned above because these solutions are designed to provide a completely hands-free interaction and greater mobility by not relying on any external switch. Instead, they require the user to hover over a target in graphical user interfaces for an activation time to select the target. As several comparative studies \cite{zuniga2017camera, magee2015camera} show, dwelling method is quite slower than clicking method for selection. It also has the Midas Touch problem, unintentional selection due to `pointing-hold' on a random target and this issue limits the design graphical user interface since bigger targets are necessary \cite{penkar2012designing}. 

\subsubsection{Pointing Methods:}
Adaptive Pointing Devices are several adaptive mouse solutions with different shapes and sizes to provide easier use in the market in addition to larger touchpad and touchscreens that recognize customized gestures like Mott et al. \cite{mott2016smart} suggest. While these solutions provide greater accessibility than standard input devices, they still require fine physical ability that majority of people with motor impairments do not have. Also, having an external device or an extra large touchscreen counteracts the logic of complete mobility that our approach brings.

\textit{Screen Scanning mechanisms} scan in a specific order along horizontal and vertical axes of the screen in a loop until the user selects the desired target by a switch. Although this is the only practical interaction technique when the user has no physical ability, it functions as a single switch only in the worst cases. This is the least efficient pointing method in terms of time since the system needs to scan two dimensional targets in one dimension and the cost of missing item selection during the scan is the twice of looping time which would be very long when the number of items is considerably large. Like most of today's popular operating systems have, iOS also has a built-in scanning mechanism, Switch Control, \cite{switchcontrol} that supports both item and point scanning. 

\textit{Gaze-tracking:} The gaze-based pointing methods calculate an approximate gaze point by tracking the movement of eye components where the user has no other physical ability. n these unfortunate cases, gaze-based techniques \cite{majaranta2011gaze, kurauchi2016eyeswipe, zhang2017smartphone} connect the user to the outside world by building a direct communication. There are also other gaze-based solutions with more sophisticated interaction \cite{tobiidynavox, switchcontrol, eyecontrol, mygaze} that allows users to play games and browse the web. But these solutions generally require to use a well-calibrated external device under a fixed lighting condition. Also, comparisons between head-based and gaze-based interactions \cite{bates2003eye, kyto2018pinpointing} conclude that head-based techniques are more voluntary, stable and have greater accuracy while gaze-based techniques would be faster for some specific tasks like typing \cite{gizatdinova2012comparison}. Despite experimental studies \cite{huang2015tabletgaze, krafka2016eye, ranjan2018light} that have potential to increase its accuracy on mobile platforms, the practical mobile usage of gaze-based interactions are still limited.

\textit{Head-based Pointing:}
Head-mounted stylus has a long history. It is used as a writing tool by attaching a regular pencil at the edge of the stylus. Today, we still have physical head-mounted styluses \cite{polacek2013nosetapping} for touchscreens and sophisticated products like Quha Zono \cite{quhazono} and Glassouse \cite{glassouse}. Besides physical devices, an important number of today's alternative pointing methods employ visual-based interactions \cite{manresa2010user} that detect and track a voluntary movement of a body part \cite{roig2016evaluation} for two-dimensional pointing. Betke et al. \cite{betke2002camera} show that visual tracking of body features, especially facial [e.g. face, nose, eyebrow], can be a successful pointing tool for people with motor impairments. Mauri et al. \cite{mauri2006computer} also review assistive technologies for the same user group and conclude that visual-based systems would be the only way of computer interaction for some users. Advantages include flexibility and lower cost over other traditional assistive technologies. Today, there are many successful applications of head-based pointing in assistive technology (e.g. Camera Mouse \cite{cameramouse}, Smyle Mouse \cite{smylemouse}, Enable Viacam \cite{viacam}, HeadMouse Nano \cite{headmouse}, EVA Facial Mouse \cite{evafacialmouse}). In addition to assistive technology, we would also like to highlight the huge potential of head-based tracking in other areas, including desktop GUIs \cite{bichsel1993automatic}, wearable computing \cite{brewster2003multimodal}, and VR 3D user interface \cite{bowman20043d, clifford2017jedi, kyto2018pinpointing}.  

\textit{Evaluation of head-tracking on mobile:}
Today, there are also ready-to-use mobile head-tracking solutions such as EVA Facial Mouse \cite{evafacialmouse} and Essential Accessibility \cite{essentialaccessibility} available on mobile platforms that rely on head-tracking and provide considerably free control of the mobile environment without any external devices, nor requiring a sensitive calibration. But these kind of products are only available on Android. On the other hand, we find that recent studies \cite{roig2018head, roig2016evaluation,roig2017evaluating, abbaszadegan2018trackmaze} that evaluate head-tracking on mobile devices tend to use iOS as their experimental environments. In their first study, Roig-Maim{\'o} et al. \cite{roig2016evaluation} propose two similar tasks to evaluate head-tracking, a picture-revealing puzzle game for pointing and a item selection task for different sized items. In their most recent work,  Roig-Maim{\'o} et al.  \cite{roig2018head} applies an user performance evaluation through Fitts' law for a mobile head-tracking interface by following the multidirectional tapping test described in the ISO standard \cite{iso9241} after his non-ISO study \cite{roig2017evaluating}. To the best of our knowledge, these are the first and still only Fitts' law studies on user performance of head-tracking interfaces with mobile devices.


\subsection{Augmented Reality}
Augmented Reality (AR) is another exciting field and there are several recent developments \cite{barfield2015fundamentals, billinghurst2015survey} that potentially transform the way people play games \cite{lv2015touch, thomas2012survey} or experience online shopping \cite{ speicher2018virtual, yim2017augmented}. The most popular mobile platforms (e.g. iOS and Android) encourage their developers to build more integrated applications with AR by providing ready-to-use native AR development kits (e.g. ARKit \cite{arkit}, ARCore \cite{arcore}). Smart phone manufacturers like Apple upgraded the features (i.e. TrueDepth Camera \cite{truedepth}) of their products to support AR applications. While these exciting advancements are happening right now, we have discovered a smart way to benefit from these recent improvements for everyone. The augmented facial expressions/masks are one of the most common applications of AR and rely on the sophisticated Computer Vision tasks such as precise head pose tracking and facial gesture detection \cite{canessa2014calibrated, chen2015augmented} including gaze. Recently, Apple released iPhone X with its front TrueDepth Camera \cite{truedepth} to support further advanced AR features available in the ARKit 2 \cite{arkit}. This toolkit serves as black-box on complicated head pose and face tracking and provides precise localization in 3D space. AR developers can easily access these measurements in few lines of code. In our work, we utilize ARKit 2 to develop an Assistive Technology that relies on the same measurements and project from real world coordinates to screen coordinates.

\subsection{Online Shopping Interfaces}
Online shopping has been growing incredibly fast \cite{Census2017quarterly} and keep expanding its tentacles globally ever since the dot com boom in the 1990's \cite{turban2017electronic}. Nowadays, more and more consumers choose to go shopping online instead of in-store as it allows them to shop 24/7 from anywhere and to compare the products more easily among a greater varieties \cite{kpmg2017truth}. Monsuwe et al. \cite{perea2004drives} also points out additional factors that encourage consumers to shop online. These include situational factors, product characteristics, previous online shopping experiences, and trust in online shopping besides ease of its use. 

For both online and in-store shopping, consumers need to search for a specific product among many other choices, choose the desired ones and compare them until final checkout. Any online shopping service needs the functionality of showing multiple products and allowing the user to select desired choices. Since the number of listing products cannot fit the screen size of devices in most cases, almost any Graphical User Interface (GUI) design for online shopping relies on scrolling to allow the user browse enough number of products.
\section{Design Process}


\subsection{Initial Observations}
These were our observations prior to start of this work. Firstly, people with motor impairments have several difficulties during shopping. We overview the literature and discuss our findings in the \textit{Introduction} section.Current interface of online shopping application has its own barriers for people with motor impairments since it heavily relies on standard interaction methods. We highlighted this limitation in the related work subsection. Besides, standard interaction methods (i.e. selection, pointing and typing) need to be carefully adapted for people with motor impairments. Although sophisticated interactions introduced to  Virtual and Augmented Reality inspired our work, they are nothing short of limitation in the mobile environment.
We surveyed relevant interaction methods individually in corresponding subsections under \textit{Related Work} and try to choose the most optimal solution. Among alternative interaction methods for people with motor impairments, there are only a few that provide true mobility. This is especially true on iOS platform where open-source tool for hands-free pointing is sparse. Although Apple ARKit is mainly intended for entertainment e.g. Pokemon Go and Animoji, we saw the potential of taking advantage of its head/face tracking capability to empower mobile interface with hands-free option.

\subsection{Design Requirements}
%
%
%
%
The aforementioned observations and the urgent need to enable motor-impaired person using mobile device with minimum effort inspired us to create a mobile hands-free control without any auxiliary hardware that can be easily to integrated into existing applications. To be specific:

\begin{itemize} 
\item \textit{Hands-free} is a must-have as we want to empower the conventional mobile applications to be accessible to everyone including motor-impaired persons while reducing the friction in target-pointing and -selection. 
\item \textit{Auxiliary-free} 
Any external device or sensor brings additional cost and undermines mobility in most cases.
\item \textit{Mobile-friendly} Nowadays, smartphones are ubiquitous and has become the main source of information we consume everyday. Being mobile-friendly means we can maximize the benefits of our heads-free control on the most common platform and be influential in numerous existing and future mobile applications.
%
%
%
%

\end{itemize} 
 
\begin{figure*}[ht]
\centering
\begin{tabular}{c}
    \includegraphics[width=0.65\linewidth]{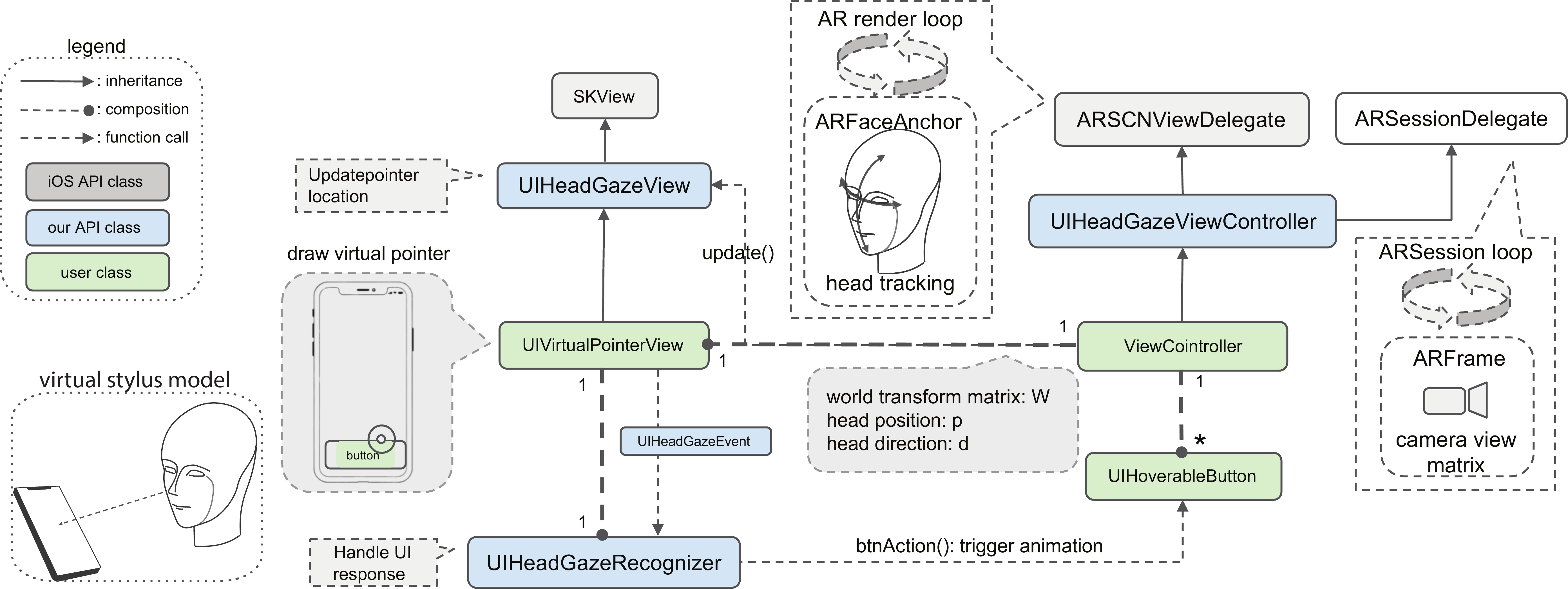}
\end{tabular}
\caption{Mobile head tracking architecture to control button using head ``pointing''. We created a set of modules to extend native UIKit to handle head tracking, pointer visualization, head gaze event detection, triggering and forwarding.}
\label{fig:flowchart}
\end{figure*}

\subsection{Design Decisions}
Next, we make the following design decisions to meet each requirement in the most compact way. 

\subsubsection{Interaction Methods}
We propose a \textit{Completely mobile} solution to the defined problem. 
After comprehensive literature search, we concluded that people with motor impairments have serious difficulty to use regular mobile devices and that the solutions rely on external devices, thus limiting mobility scientifically. To make our solution \textit{hands-free} and \textit{auxiliary-free}, we evaluate all possible methods for human-computer interaction in the \textit{Related Work} section. These comparisons led us to pick head-based \textit{pointing} with dwelling function for \textit{selection}. With this combination, we make sure our solution relies only on the embedded features of the device - specifically the front camera, since the methods we prefer are visual-based interaction methods. 

\subsubsection{Application for Online Shopping}
We would like to enable people with motor impairments to be able to shop online on smartphone and feel independent. We design an online shopping application that requires minimum effort to interact with. To keep our design even simpler, we build only browse-based shopping application that allows users to browse products across categories. This also eliminates other complex tasks such as searching and ranking which are required for a regular online shopping application. While a  browse-only shopping application does not provide a complete shopping experience, it is still important since our approach will be one of the very first assistive eCommerce applications. Also, since we open source this project, anyone could extend its functionality by adding searching logic on top of our work. 

\subsubsection{Development Environment}
With the intention of developing a mobile online shopping that only depends on visual-based interaction, head-tracking specifically, we searched for the most appropriated environment that has the necessary functionalities. Instant head-tracking is still a challenging task on mobile devices since it relies on power-hungry video processing and ability to capture high quality video through front-facing camera. This solution is not practical on many mobile device with limited resource and lead us to focus on the most recent product releases only. iPhone X has TrueDepth Camera \cite{truedepth}, an impressive feature. It was originally designed to support advanced AR applications on iOS 11 and accessible through Apple's ARKit 2 \cite{arkit}, an AR development SDK. ARKit 2 implements efficient on-device head-tracking and exposes precise measurements of several facial landmarks to the developer. The biggest advantage of this setup is that it only relies on the device's built-in components without requiring extra accessories. This is power efficient and portable. 
Our open-source head-based interface components on iOS were built on top of ARKit 2. Even though the tools we built exist only on the most recent iPhone model, which may limit its usability, according to the report of Strategy Analytics \cite{StrategyAnalytics2018iPhoneX}, iPhone X is the world's best-selling smart phone model, shipping an impressive 16 million units during the first quarter of 2018. These numbers together with the promising future of AR applications lead us to believe that most of the smartphones on the market would intend to adopt similar features in the near future. Overall, a smart phone based solution is more affordable, portable, and more beneficial than many of the common external solutions that heavily rely on separate gaze or head-tracking gadget.

\section{System Design and Implementation}
\subsection{Head-based Pointing}
To meet our design requirements, we first implement a \textit{hands-free}, \textit{auxiliary-free}, \textit{mobile-friendly} interaction system. Our system consists of two main modules. A module for head-based pointing, and a customized UI module for the interaction with on-screen pointer and feedback visualization.


\subsubsection{Head-tracking}
ARKit 2 comes with stable head-tracking capability where the 3D location, orientation of the tracked head can be queried at each AR session through \textit{ARFaceAnchor} object. 
Abbaszadegan et al. \cite{abbaszadegan2018trackmaze} used the same technique to highlight the potential of using the TrueDepth front-facing camera of an iPhone X for tracking.

\subsubsection{Head-to-pointer Mapping}
However, the limitation of ARKit head-tracking is that it does not provide additional information regarding the location on the screen at which the head or nose is pointing. To bridge this gap between 3D head posture and pointing location on the phone screen, we designed an intuitive virtual stylus model to give user stable control of their pointing direction and instant feedback of the pointing location on screen. Figure \ref{fig:flowchart} illustrates how the model works. To aid calculation of pointing location, we model the phone screen as a 2D plane perpendicular to the z-axis at position z=0. A proxy ray is spawned from the head center and passes through the nose. The ray-plane intersection is then calculated and is considered as the pointing location. Another possible solution is to project nose position directly on the screen and take its screen space coordinates as pointing location. We found that in practice the virtual stylus model takes into consideration the distance between head and phone, which allows distance-based sensitivity adjustment. Such feature is adorable as it adapts to UI interface that features widgets/controls that vary in size, shape, and gap.
To compute the intersection, let $\mathbf{p}_{0}\in\mathbb{R}^4$ denote the homogeneous coordinates of the original head position in the object space, and $\mathbf{W}\in\mathbb{R}^{4\times4}$ denote the world transformation matrix from \textit{ARFaceAnchor} object updated at each frame. The initial virtual stylus direction is denoted by $\mathbf{d}_{0}\in\mathbb{R}^3$ and points to the negative $z$ direction perpendicular to the screen. At each frame, the head center is updated to $\mathbf{p}=\mathbf{W}\cdot \mathbf{p}_{0}$ and the pointing direction is updated to $\mathbf{d}=\mathbf{W}\cdot \mathbf{d}_{0}$. The resulting ray-plane intersection $\mathbf{b}$ in Normalized Device Coordinate (NDC) space $[0,1]\times[0,1]$ is then calculated by $\mathbf{b}=\mathbf{p} + \mathbf{d}\times t$, where $t=-\mathbf{p}.\mathbf{z}/\mathbf{d}.\mathbf{z}$. We further apply viewport transformation to map the NDC coordinates to the screen space that can be used for UI widget intersection test.

\subsubsection{Pointer Visualization}
To give user instant visual feedback on head-pointing, we visualize the location as an on-screen cross-hair using Apple SpriteKit API. Specifically, to cover the full screen and to provide precise tracking feedback, we add to our application a \textit{UIHeadGazeView} layer that inherits from \textit{SKView} to make \textit{UIHeadGazeView} a Spritekit Scene object that serves as a canvas on rendering 2D geometry on the screen to represent pointing location.


\begin{figure*}[!ht]
\centering
\begin{tabular}{ccccccccc}
    \includegraphics[width=0.1\linewidth]{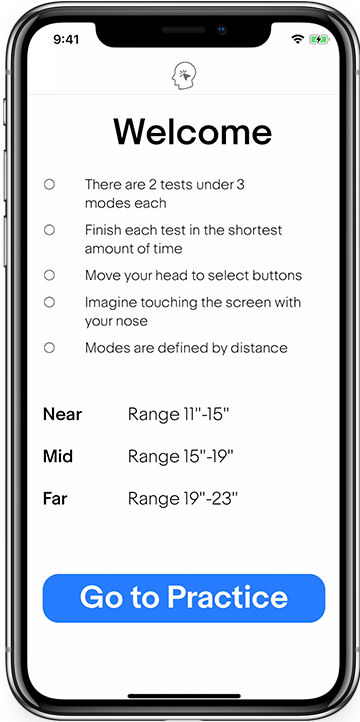} & \hfill &    \includegraphics[width=0.1\linewidth]{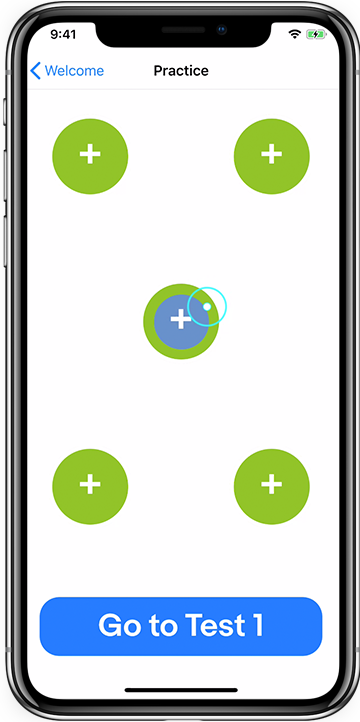} & \hfill &    \includegraphics[width=0.1\linewidth]{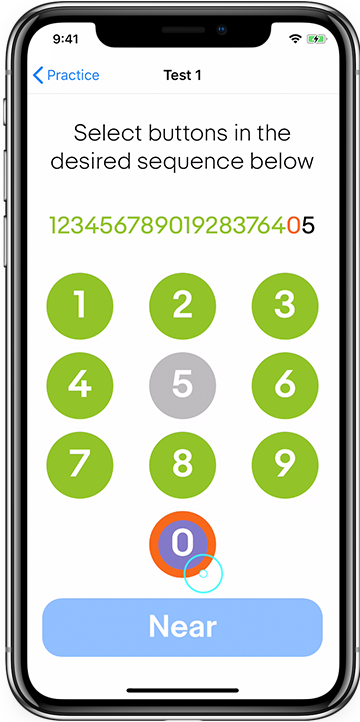}& \hfill &    \includegraphics[width=0.1\linewidth]{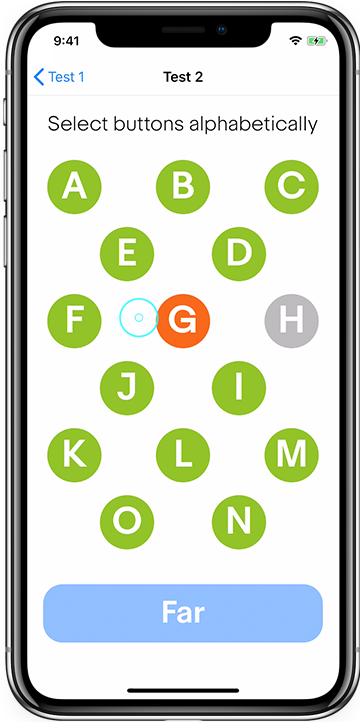} & \hfill &    \includegraphics[width=0.1\linewidth]{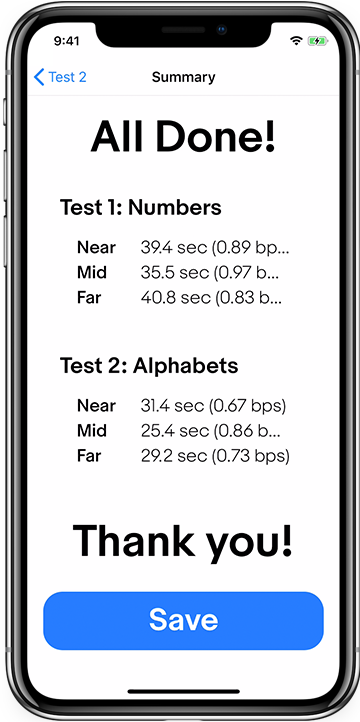} \\
    (a) & & (b)  & & (c) & & (d) & & (e)
\end{tabular}
\caption{UI design for user study via 2 tests. Each test is performed at 3 different distances between face and screen (near, mid, far). A test is complete when the desired sequence is complete. Next button in the sequence to be selected is shown in orange. To reduce response time, we highlight the following button in the sequence in gray. Buttons fill up blue during dwell time after which the next button is highlighted in orange and the following button in gray. (a) welcome screen with instructions (b) unlimited practice session for warm up in different distance modes (c) first test for densely packed button around the center of screen (d) second test with buttons uniformly distributed across the screen (e) summary is shown in the final screen.}
\label{fig:sensitivity_screenshots}
\end{figure*}

\subsection{UI Module}
\subsubsection{Selection Method}


Native UI widgets implemented by the Apple UIKit package only react to user physical touch event such as touch down, touch up, pinch, etc. Unfortunately, there is nothing to support interacting with the UI from mid-air. This requires a new definition of interaction for head-based pointer and native UI components. To bridge this gap between the pointing location and interaction with existing UI, we defined several interactions and implemented them through a set of customized UI widgets on top of Apple UIKit so that they can be seamlessly integrated into new iOS applications or extend the existing one with minimum effort. For example,
in Swift, we created a \textit{UIHoverableButton} class by extending iOS' default \textit{UIButton} class to respond to the pointer interaction. Such customized button senses the pointer's location change and triggers hovering events whenever the pointer enters the button, and will further trigger selection event if the pointer has been hovering beyond a user specified time interval, hence dwelling. During pointer hovering, a user customized animation effect can be added to the \textit{UIHoverableButton} to serve as visual cue on the elapsed hovering duration. For instance, by gradually increasing the button size or gradually filling the background with different colors. See Figure \ref{fig:Teaser} for illustration.

Besides \textit{UIHoverableButton}, there are other customized classes implemented to extend native UIKit to support head tracking, pointer visualization, head gaze event detection, triggering and forwarding. The inter-module connection is illustrated in Figure \ref{fig:flowchart}. At high level,  \textit{UIHeadGazeViewController} keeps track of head motion and passes the head's world transformation matrix to \textit{UIHeadGazeView}. \textit{UIHeadGazeView} then computes the head-to-pointer mapping, updates current pointer location, and notifies the registered event recognizer \textit{UIHeadGazeRecognizer} about the updates. These updates are encapsulated in a \textit{UIHeadGazeEvent} object. The design follows iOS UIKit protocol and blends into regular event handling scheme. Such design requires minimum effort to modify existing code base to support our head-based pointing feature.

User can select different types of \textit{UIHeadGazeEvent} depending on how long the pointer needs to hover over a button before triggering the corresponding type of \textit{UIHeadGazeEvent}. We currently have predefined two types based on the length of the dwelling duration: glance (1 sec) and gaze (2 secs). Trade-off has to be made on setting the duration, as longer duration reduces the chance of accidental clicking but increases latency. By default the duration is 1 second (glance). This configuration reduces the amount of accidental clicking of unintended buttons while maintaining the interaction at high throughput.
Furthermore, to avoid unintentional repeated clicks on the same buttons, the button is configured not to emit another \textit{UIHeadGazeEvent} until the pointer reenters that button. With this limitation, the user has to move the pointer away from the button to emulate touch up before they can re-click the button if needed. While this is an extra effort for intentional re-clicking, the effort is actually negligible compared to the one incurred by undoing the unintentional re-clicking, which normally requires clicking different buttons to reverse it. 

\subsection{Simple Online Shopping Interface Design}
Once we completed the implementation of our head-based interaction system, we integrated this system into a simple shopping application to allow motor impaired person to shop online.

\subsubsection{Product Browsing Page}
Our shopping application starts with a product browsing page which takes up the main screen of the application (figure \ref{fig:Teaser}). We present  one product at a time on the screen with a reasonably large photo along with title and price information. At the bottom of the screen are the three action buttons which allow users to share item on social media, checkout item, or add item to shopping cart. Each of the three action buttons brings up a separate hands-free page for the subsequent process. For the interest of space, we focus our discussion on the main product view page.
Items are organized in a 2D grid and grouped in each row by their category. The item view is surrounded by four direction buttons that allow users to browse through items (horizontal swipe) and categories (vertical swipe). Users are presented with one cell of the grid at a time while they can swipe the item via "clicking" the direction buttons. 

\subsubsection{Open Source}
We open-sourced the head-based pointing solution with advanced mobile UI components compatible with head-based pointing. We believe such generic library can benefit both researchers and developers in many different use cases. Through a sample online shopping application, we addressed several design concerns and showed practicality of our solution.

\section{Experiments}
Beyond the online shopping application we implement, we believe that the primary contribution of this work is the open-source tools for head-based pointing on iOS which is one of the most common mobile platforms all around the World. To see the first user experience and understand the relative performance, usability, and user preference of these tools, we conducted a field experiment with 75 able-bodied participants during an internal showcase of our organization. Despite we state our target user group as people with motor impairments, anyone would benefit from a hands-free interaction with smart phones. Notice that the physical abilities of people with motor impairments varies a lot from person to person and the assistive technology solutions they currently use would also vary. In this case, there is high possibility of unfair evaluation of the proposed system by only limited number of people with motor impairments. We believe that providing a baseline evaluation of this system with able-bodied participants would be an important contribution since the results would inspire the community for the future works. In addition to the field study, we also recruited 27 able-bodied participants for a lab-based user study to test our head-based interface for possible future work not limited by online shopping. Finally, We conducted a simpler experiment by hiring a person (24, male) with motor impairments to try the same interface and provide his feedback for comparison.

\subsection{Participants}
We conducted 2 separate experiments. In the field study, we recruited 75 able-bodied unpaid participants (35 female, 40 male) to experience our online shopping app. We randomly picked the participants during a crowded event and solicit their initial feedback on our approach right after they played with the app. In the lab based study, we recruited 27 able-bodied unpaid participants from our organization and involved them in an instructed user study. Among them, 16 were male and 11 were female. In both experiments, the participants had no prior experience with head-based pointing methods, nor dwelling function. They were from different ages with various races. Some of them wore glasses during the experiment. With this diverse group of participants, we intended to get a realistic insight into the learnability and usability of the system from novice adopters of the technology since this methodology is uncommon, especially on mobile. These studies are insightful on the practicality of our approach to a real-world application such as an online shopping solution.     

\subsection{Design and Procedure}
For the lab-based study, We implemented a separate UI design as a single mobile application and conducted the experiment by running this App on an Apple iPhone X running iOS 11.2 with a resolution of $1125\times2436$ px and a pixel density of 458 ppi. This corresponds to a resolution of $375\times812$ Apple points (pt) which is an abstract unit that covers two pixels on retina devices. The camera was also the iPhone X's embedded 7-megapixel front-facing TrueDepth Camera \cite{truedepth}. We fixed the position of the phone by placing it on a holder in portrait mode. The participants were required to sit against the phone at 3 different distances for each test. We placed the phone on a holder which was attached to an adjustable table. Before the experiment, participants were also informed of the purpose and instructed to complete the tasks. The experiment lasted about 15 minutes per participant. 

The user study App for this experiment had 5 different screens (figure \ref{fig:sensitivity_screenshots}) to show in order during each session. Our user study App opens with a welcome screen with the instructions of the overall experiment. Then, we have an unlimited practice session that allows the participants to get familiar with head-based pointing. We have two consecutive tests with different layouts on separate pages to get different feedbacks on the performance of participants. During the experiment, we also collected the timestamps of the participant's each action and the position of the cursor for analysis. At the end of this experiment, we also asked each participant to fill a questionnaire for recording their user experience. The experiment is carried out on two layouts with three distances: Near (11 to 15 inches), Mid (15 to 19 inches), Far (19 to 23 inches).  

\textit{Test 1} (Numbers) was designed for densely packed targets around the center of screen (figure \ref{fig:sensitivity_screenshots}c). The first test screen contains 10 targets that were labeled from zero to nine by digits similar to the numeric keypads. In addition to highlighting the next target in orange by the predefined order, we also displayed the whole desired sequence on the upper part of screen and colored the sequence accordingly. The desired order was [12345678901928376405], we especially selected this sequence to catch participants' performance on vertical, horizontal and diagonal trials with different lengths.

\textit{Test 2} (Alphabets) was with targets uniformly distributed across the screen (figure \ref{fig:sensitivity_screenshots}d). We aimed to see the reachability of targets which positioned on different parts of the screen. Test 2 had 15 same-sized targets which were labeled by the letters, from A to N. The targets had almost the same distances to their neighbors and were required to be selected in the alphabetic order. Here, we expected that the participants could select the targets around the center of the screen easily while it could be hard to select the targets far from the center, especially the targets at the bottom. 

In total, there were 27 participants $\times$ 3 distances $\times$ (20 + 15 targets) = $2835$ trials in our user study.

\subsection{Results}\label{Results}
Our lab-based study tended to precisely evaluate the practicality of the proposed interaction. Therefore, we applied a quite similar study with Fitts' Law \cite{fitts1954information} since it is the standard way to evaluate this kind of approaches and derive the dependent measure throughput (Fitts' \textit{index of performance}) as part of the comparison and evaluation \cite{roig2017evaluating}.

\subsubsection{Objective Evaluation}
We calculated throughput (TP) as follows:
\begin{equation}
  TP = \frac{Effective\hspace*{1ex}index\hspace*{1ex}of\hspace*{1ex}difficulty}{Movement\hspace*{1ex}time} = \frac{ID_e}{MT},
\end{equation}

where $ID_e$ is derived from the movement amplitude $A$ and effective target width $W_e$ and $MT$ is averaged movement time per trial over a sequence. They have units "bits" and "seconds" respectively, the units of TP are "bits per second (bps)". The effective index of difficulty is also a measurement of the difficulty and user precision in completing a task:

\begin{equation}
  ID_e = \log_2(\frac{A}{W_e} + 1),
\end{equation}

where $A$ is the  movement amplitude, the distance between the centers of two consecutive targets and $W_e$ is the effective target width, calculated from the width distribution of selection coordinates made by a participant over a sequence of trials which is calculated as below:

\begin{equation}
  W_e = 4.133 \cdot S_x,
\end{equation}

\begin{figure*}[ht]
\centering
\begin{tabular}{cc}
    \scalebox{1.2}[1]
    {\includegraphics[width=0.35\textwidth]{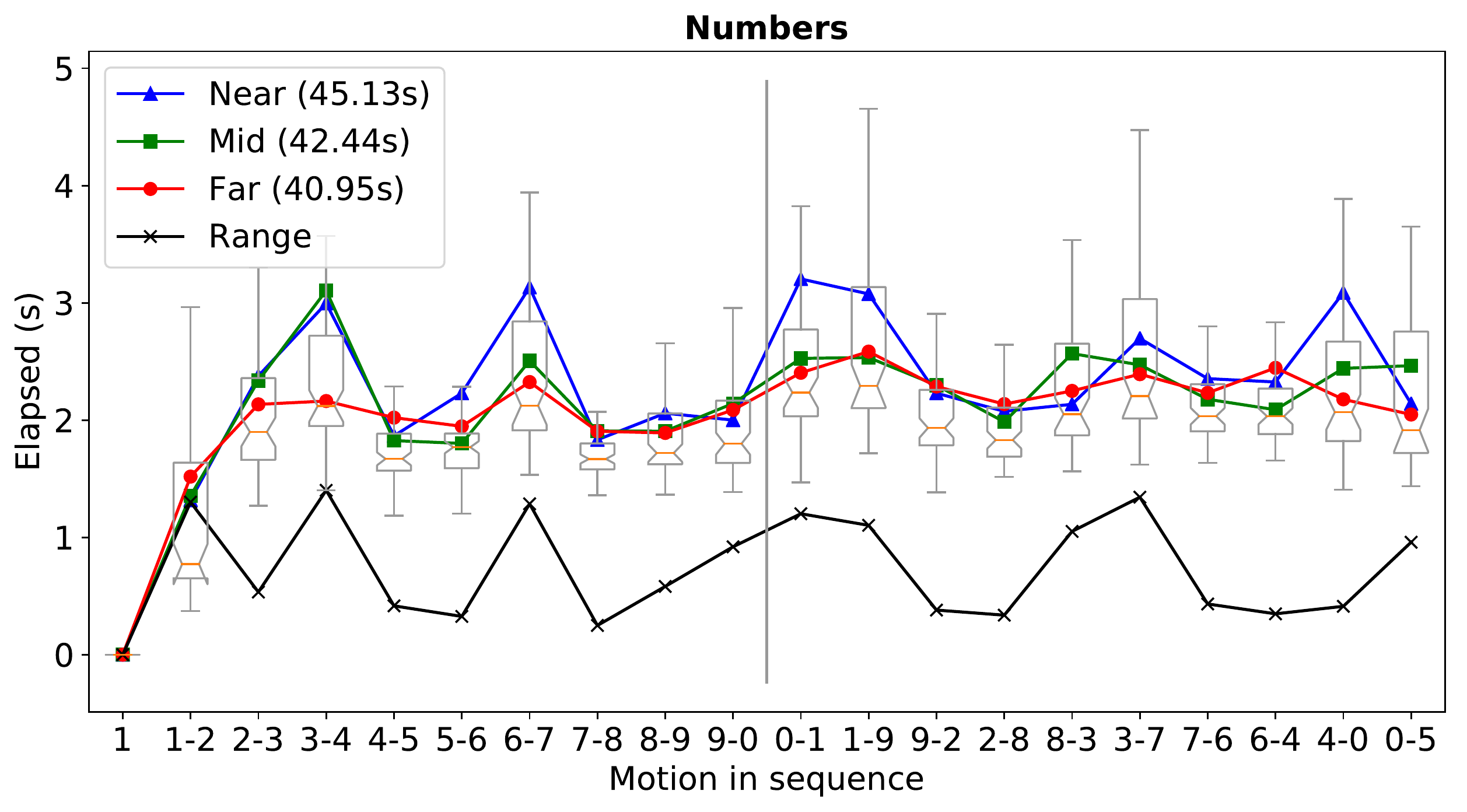}} &   
    \scalebox{1.2}[1]
    {\includegraphics[width=0.35\textwidth]{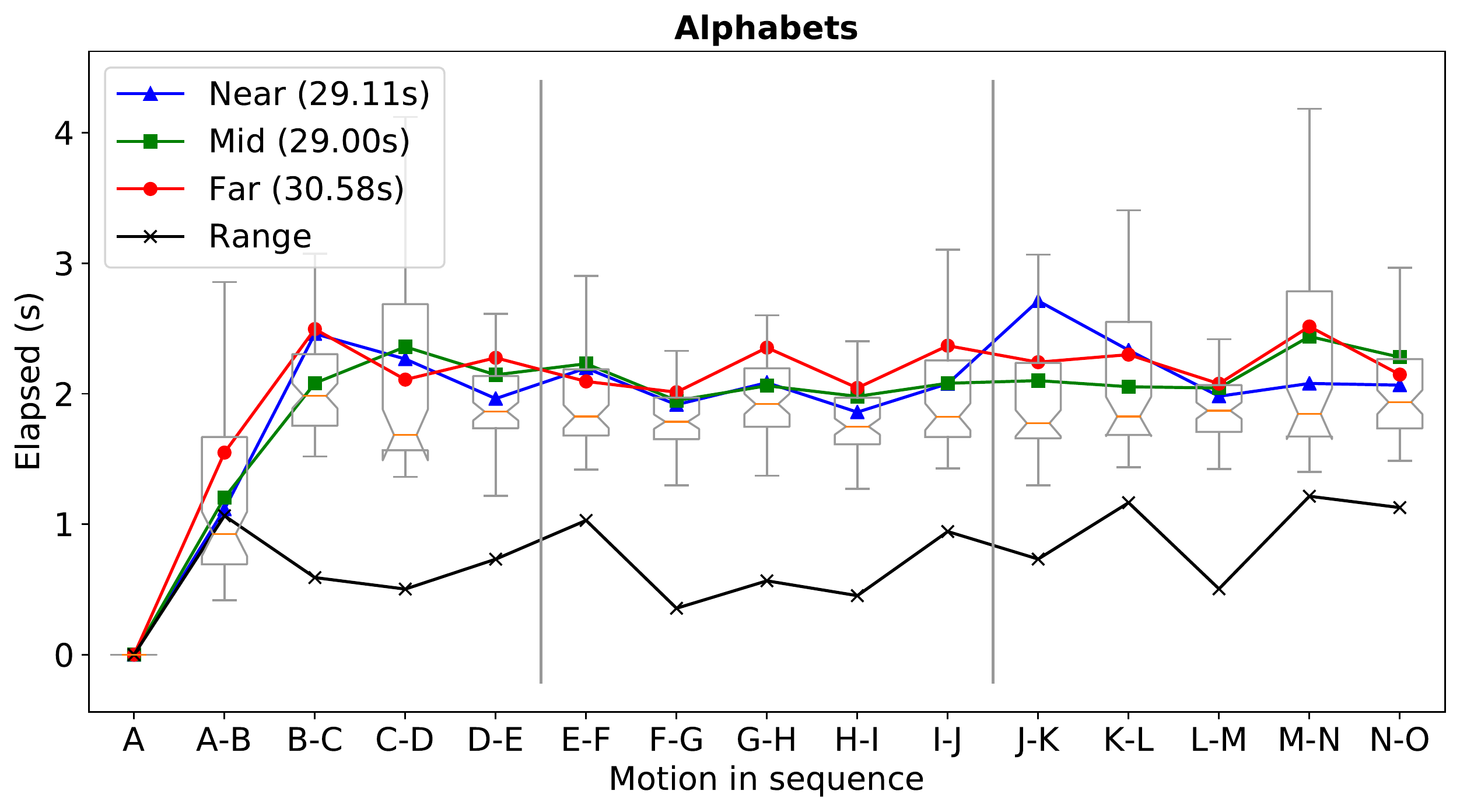}} \\
    \scalebox{1.2}[1]
    {\includegraphics[width=0.35\textwidth 	     ,trim={0 0 0 0.8cm },clip]{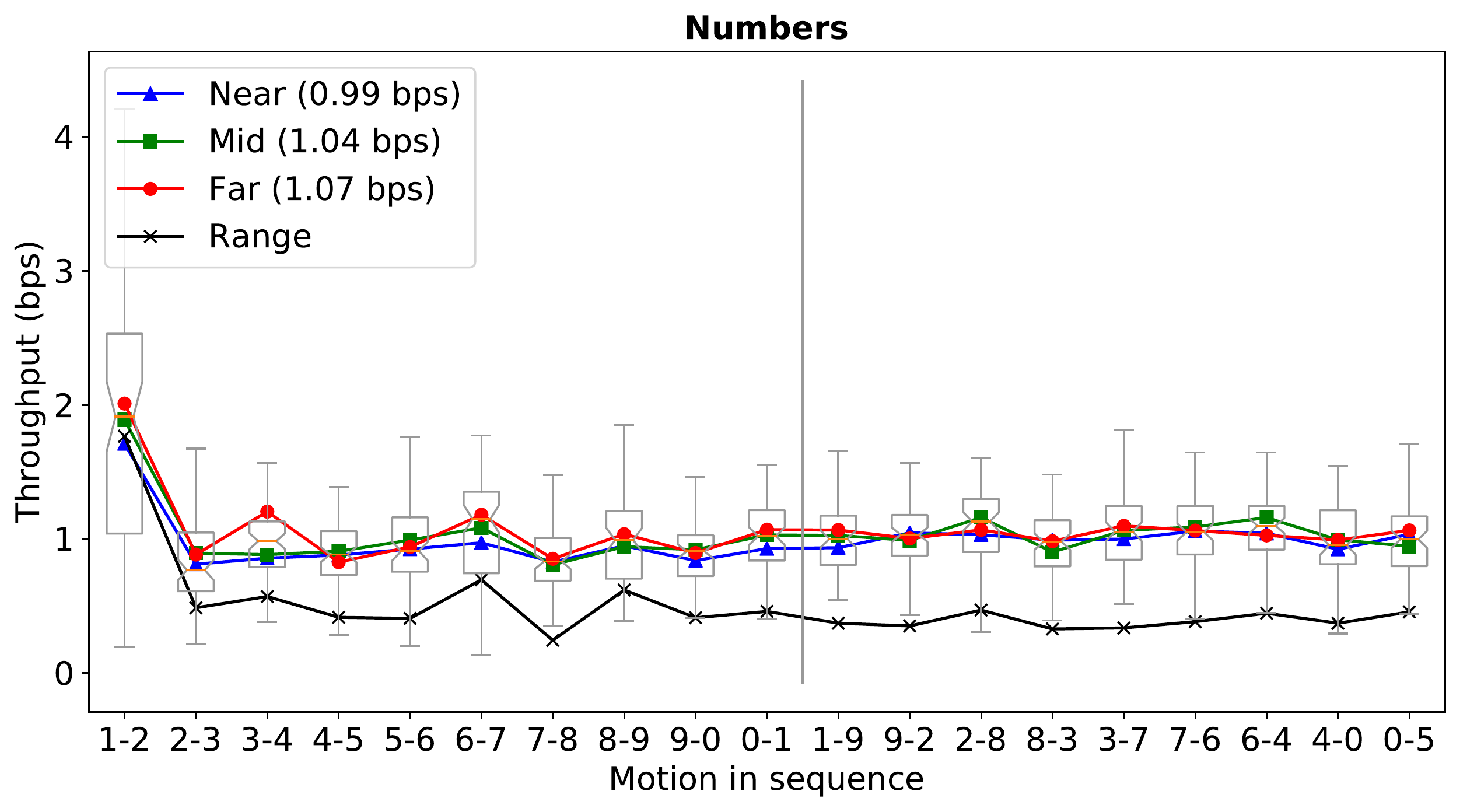}} &   
    \scalebox{1.2}[1]
    {\includegraphics[width=0.35\textwidth
    ,trim={0 0 0 0.8cm },clip]{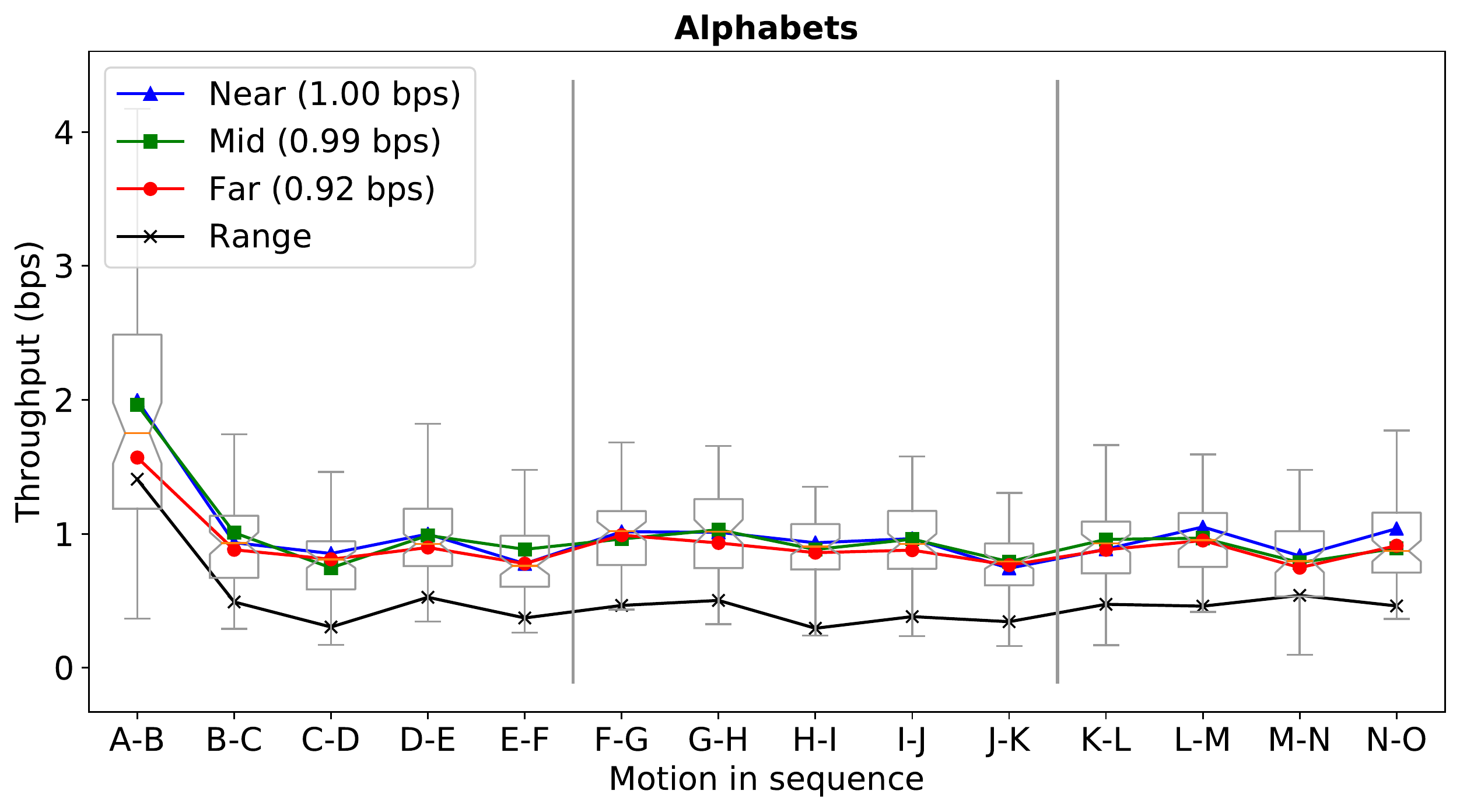}} \\
    \scalebox{1.2}[1]{\includegraphics[width=0.35\textwidth
    ,trim={0 0 0 0.8cm },clip]{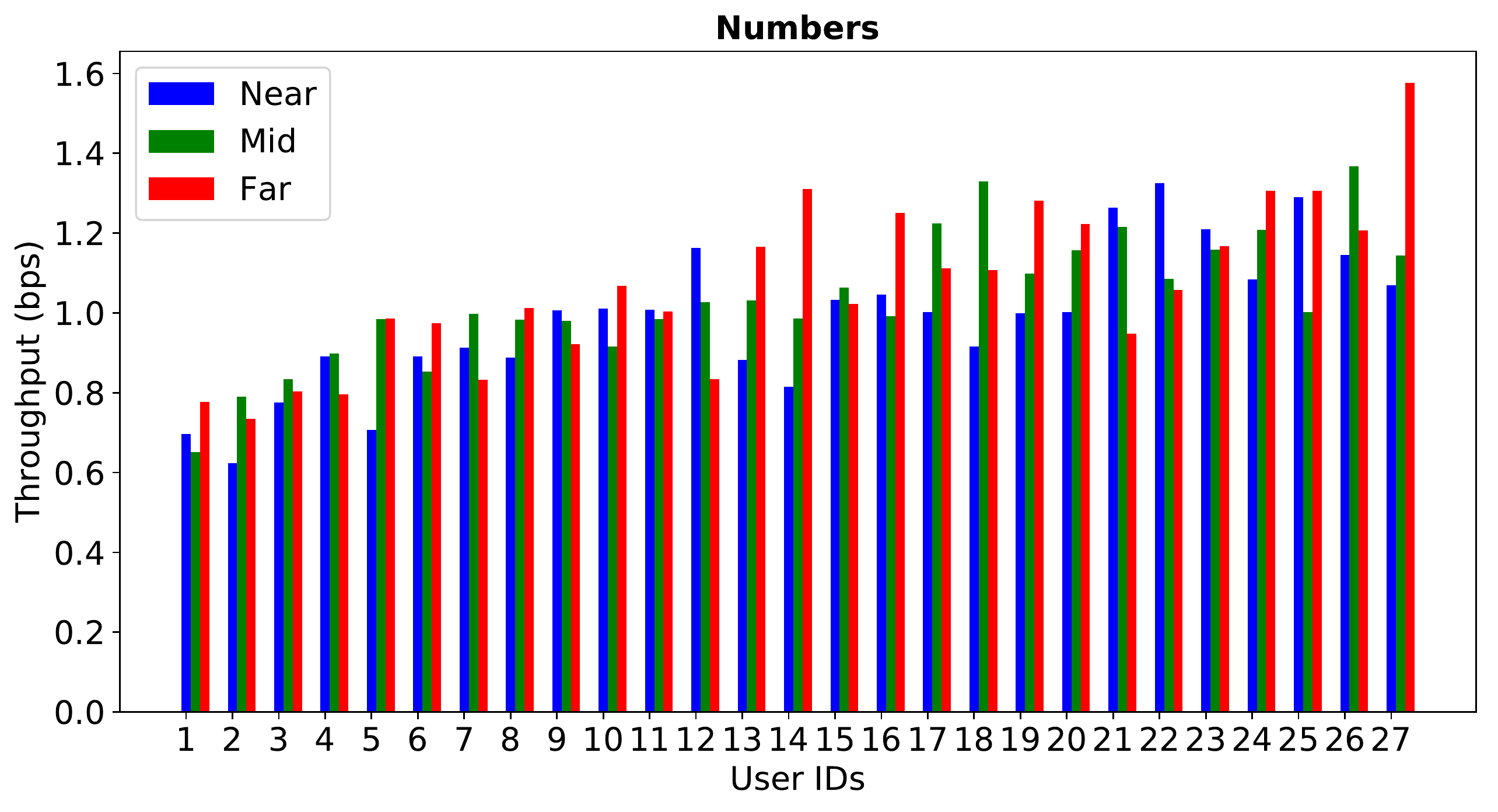}} &    
    \scalebox{1.2}[1]{\includegraphics[width=0.35\textwidth
    ,trim={0 0 0 0.8cm },clip]{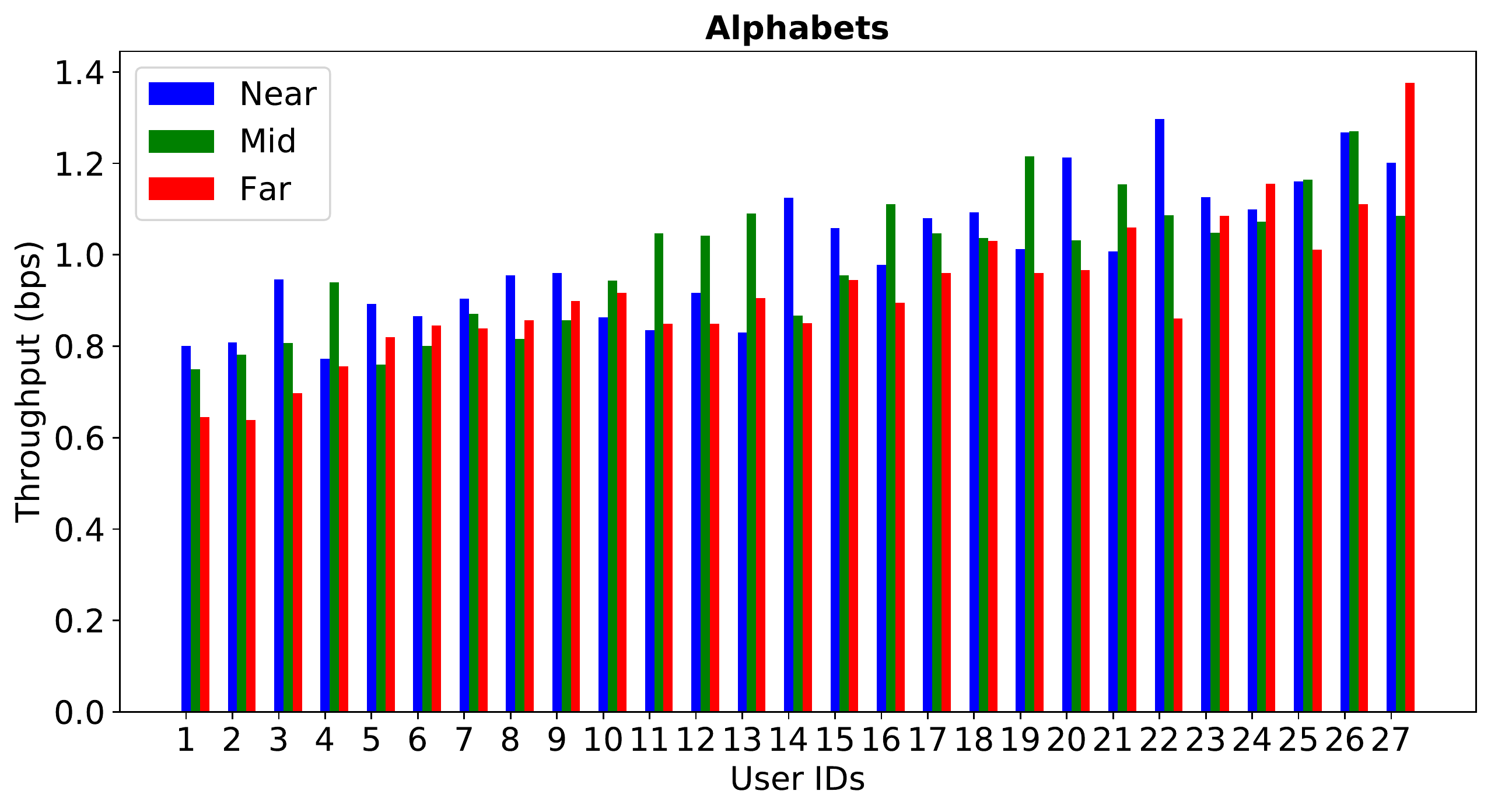}} \\
\end{tabular}
\caption{Analysis of dwell time and throughput. Top: elapsed time per motion sequence averaged over all participants. Middle: throughput per motion sequence averaged over all participants. We also show box and whisker plot for data collected from all participants at each sequence. The range plot shows median of the range of a set of data collected from all participants across all 3 distances. Bottom: throughput per participant averaged over all motion sequences, sorted by throughput averaged across all distances. Note that for the numbers experiment, the throughput is generally higher at far distances, while for the alphabets experiment, throughputs are usually higher at near distances. See Sec. 6 for details}
\label{fig:sensitivity_results_elapsed}
\end{figure*}

Note that we used the standard-deviation method to calculate throughput \cite{mackenzie2018fitts}. This is because our specific design utilizes dwelling function and has no \textit{error rate} \cite{zhang2007evaluating}. Also, we followed Roig-Maim{\'o}'s \cite{roig2017evaluating} equations and applied our own Non-ISO tests. A detailed description of the calculation of throughput can be found in \cite{ roig2017evaluating, zhang2007evaluating}. We show the summary result of dwell time and throughput in Fig. \ref{fig:sensitivity_results_elapsed}. The top row shows average dwell time of all users at the 3 distances. The box and whisker plot goes from the lower to upper quartile values of the dwell time data collected from all users at all 3 distances, with a line at the median. The whiskers extends from the box to indicate the data range. The median value of the data range are plotted in black, while mean values per distance are shown in the legend. We also follow the same fashion to show throughput per motion sequence in the middle row. Finally, at the bottom row we show throughput per user averaged over all motion sequences. 

\begin{figure}[ht]
\centering
\begin{tabular}{cccc}
    \includegraphics[width=0.2\linewidth]{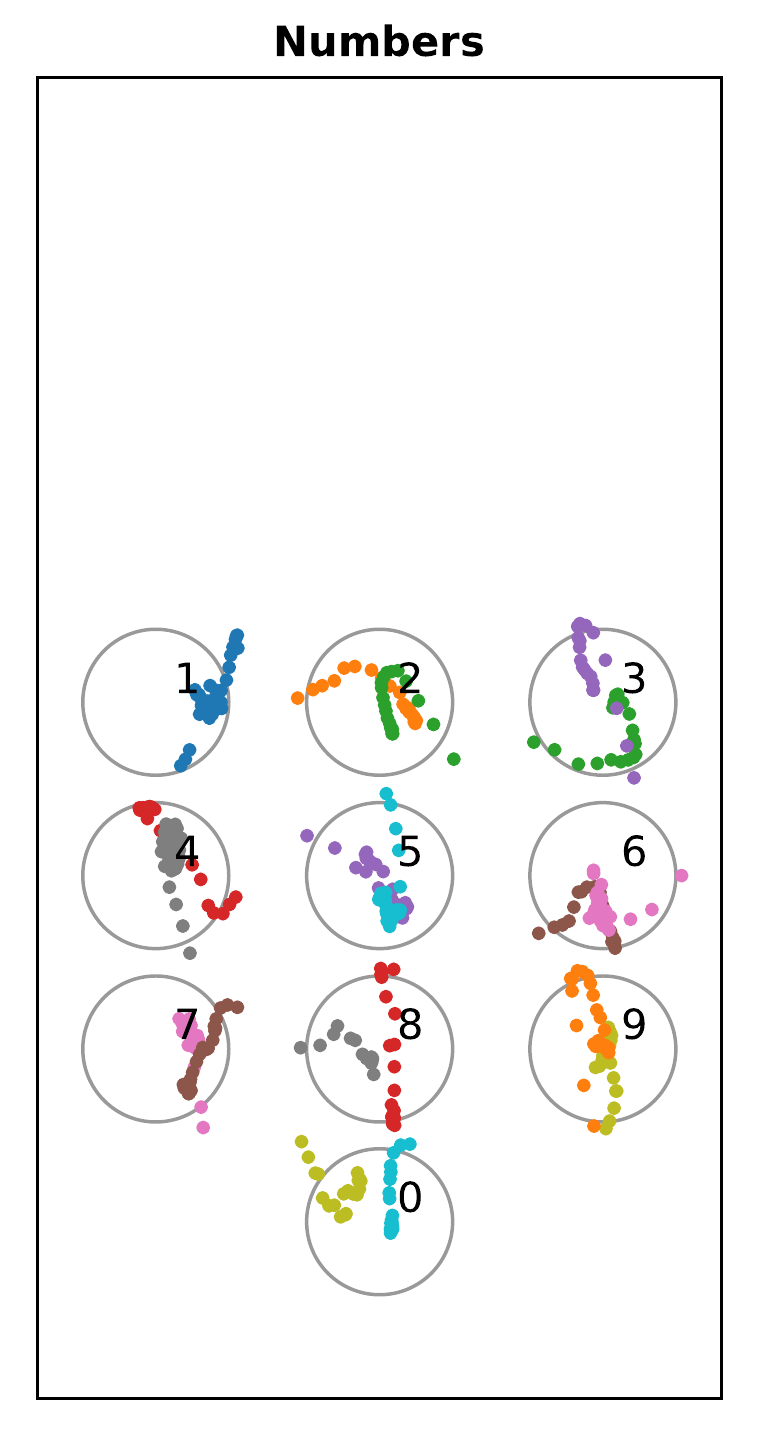} & 
    \includegraphics[width=0.2\linewidth]{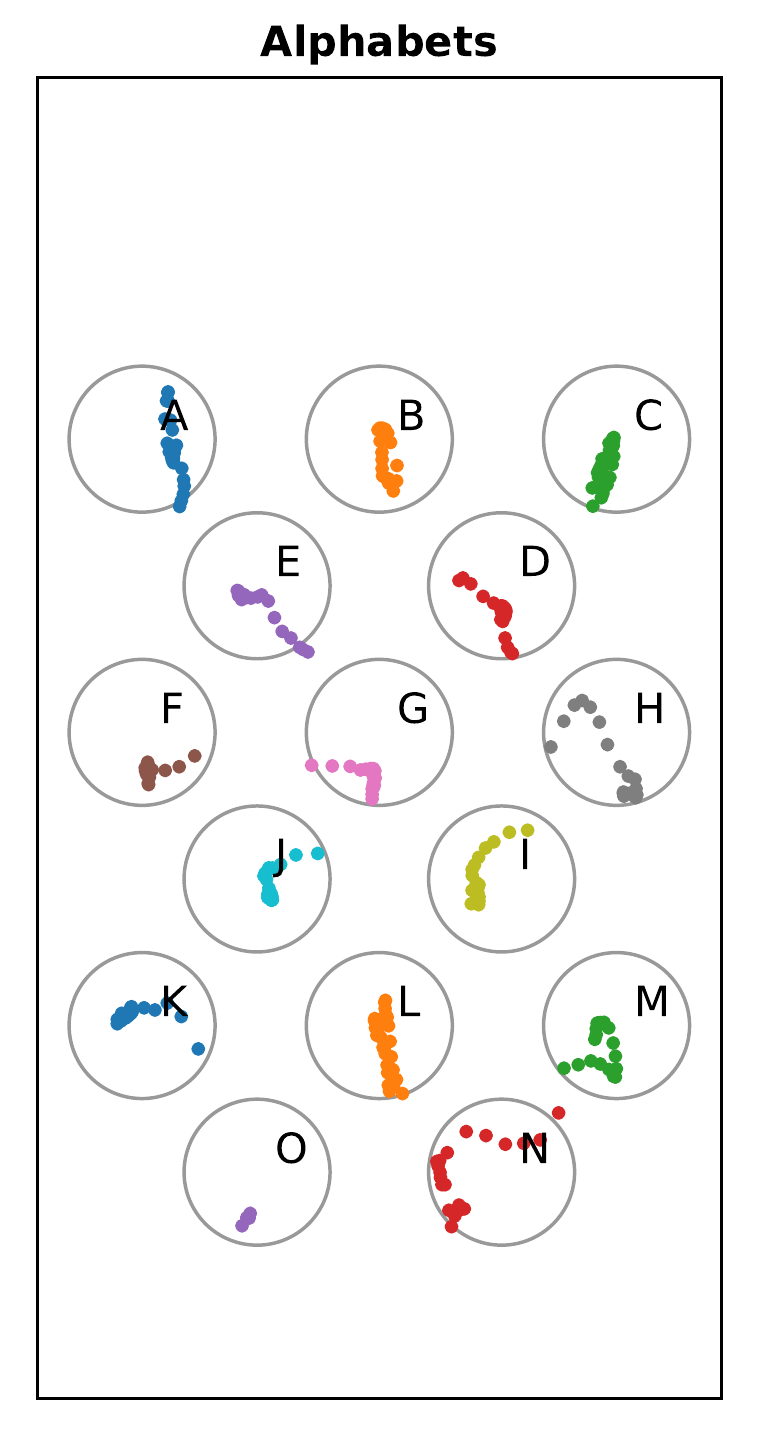} &
    \includegraphics[width=0.2\linewidth]{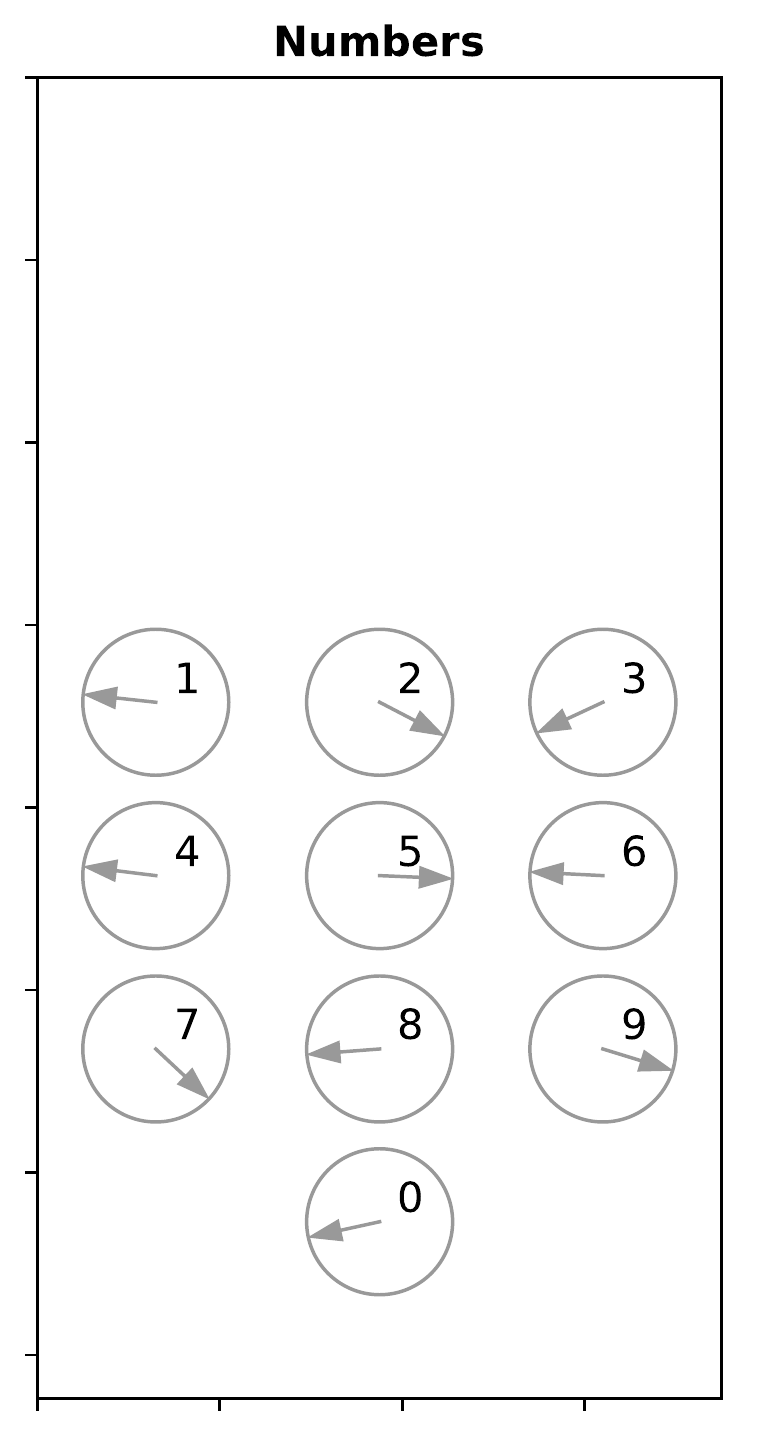} &    
    \includegraphics[width=0.2\linewidth]{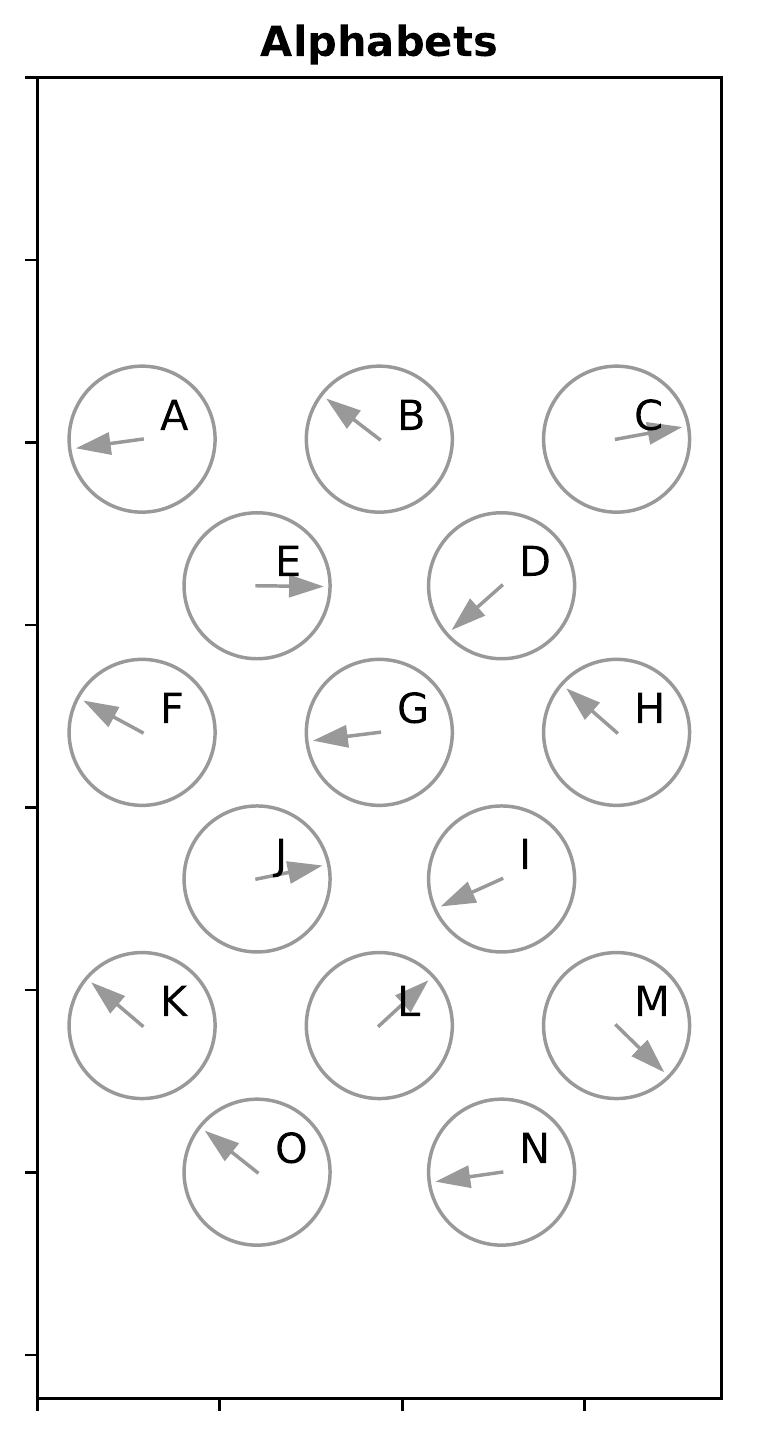} \\
    (a) & (b) & (c) & (d) 
\end{tabular}
\caption{Analysis of dwell points. (a-b) Randomly selected dwell points from a participant. (c-e) Directions of eigenvectors derived from dwell point coordinates. }
\label{fig:sensitivity_results_eigen}
\end{figure}

For the \textit{Numbers} test, it is worth noting that it is easier for the participants to navigate in the \textit{Mid} range compared to the \textit{Near} range, and most participants completed the test in the shortest time under the \textit{Far} range setting. This is because the spacing between the buttons are relatively large in this test. If the users are operating at a closer distance to the screen, it requires more head movements to control the cursor to move across larger distances. This is especially true when the participants perform large diagonal movements from number 0 to 1 or from 1 to 9. On the contrary, for the \textit{Alphabets} test, participants complete under the \textit{Near} setting in shortest amount of time with highest throughput. As the button size and spacing become smaller (See Fig. \ref{fig:sensitivity_screenshots}), it is easier to navigate in the \textit{Near} range as it offers more precise control. The above observations can also be verified at the bottom plot from Fig. \ref{fig:sensitivity_results_elapsed}. For the \textit{Numbers} experiment, throughputs are generally higher at far distances, while for the \textit{Alphabets} experiment, throughputs are usually higher at near distances. In addition to throughput calculation, we also studied the direction of selection points with respect to buttons at different locations (see Fig. \ref{fig:sensitivity_results_eigen}). We calculated the eigen-vectors of selection point coordinates on the covariance matrix to see if there is a pattern. We found that for the first column of buttons, statistics show that most users point to the left side. This may indicate that the participants are trying to position the cursor more towards the left side of the screen and most of the cursors overshoot the center position of the buttons. For the center column of the buttons, the eigen-vectors show a pattern of a path placed more towards the right side. For the third and final column, most participants tend to reverse the path more towards the left side in order to hit the buttons sequentially. On the third row, the pattern changed compared to the previous two rows, indicating participants getting more familiar with the app and thus exhibit better control.

\subsubsection{Subjective Evaluation:}
We applied a questionnaire to the participants of our lab-based study right after their individual sessions. 96.3\% of participants considered moving cursor by head was easy and normal. Majority of them also reported that the selection method was normal. 48.1\% reported there was no scientific difference between the difficulty of selection based on the target regions (upper, lower and middle portions of the screen on portrait mode). Only 18.5\% thought the lower region was hardest to reach while 29.6\% disagreed and said the targets at the upper region was hardest to select. Majority of the participants claimed practicing our tests by keeping the distance at mid from the phone was easier to use our system while 29.6\% said far distance was easier.

Our single participant with motor impairments also completed both tests and experienced the shopping App. He completed the first test slower than other participants since he had difficulty to keep his head stable for dwelling. He shows average performance on the second test as he gets more familiar with dwelling and the software interface. At the end, he found our system quite practical for several use-cases including typing. He also reported that our system's pointing accuracy is quite better than the software he uses on a daily basis for communication which runs on Windows via a web-cam. 

We also applied a questionnaire to the participants who experienced our online shopping interface during the field study. We sought their initial thoughts about the proposed hands-free online shopping App. Among 75 participants, 51 of them thought this is a useful idea. 41.3\% reported our App was easy to use even though they were not familiar with this interaction method. 31 participants also found it enjoyable while 5 of the participants could not find a use case and said this solution is not for them. Only 9.3\% claimed moving the cursor by head was hard for them while 52\% thought it was easy and 38.7\% thought it was normal to practice. Furthermore, 50 participants reported that this solution would fit somebody's needs they know.

\subsection{Discussion}
We saw that online shopping has several barriers for people with motor impairments. Our design led us to address this problem with a \textit{hands-free}, \textit{auxiliary-free}, \textit{completely mobile} \textit{online shopping sample application}. Our application allows users to browse several deals from different categories and buy/share the desired products. Users do not need any external device, nor a fine calibration step for using our approach. This eliminates the need of touch screen which requires fine motor control. The initial user feedback also shows that this interaction with an online shopping application is practical and enjoyable. Beyond this application, we actually proposed our \textit{hands-free} interface components as an open-source solution so that developers may incorporate head tracking into their mobile applications, not limited to eCommerce. Our open-source solution can be used in several applications from communication tools for people with motor impairments to a hands-free recipe or DYI application.

Our lab-based study suggests keeping the phone at mid-distance (15 to 19 inches) from the face increases the interaction quality and is more comfortable since it provides an accurate and robust interaction. There is a trade-off between cursor speed and stableness. To be specific, small phone-to-head distance requires bigger head movements that can easily cause fatigue but delivers fine control with higher stableness, while larger distance requires minor head movements but with higher precision and stable head movements which is potentially stressful as well. Future work on head-based interaction may consider this fact and design its interface accordingly. Another fact is that, while majority of the participants reported that there was no scientific difference between the difficulty of selection based on the target regions, we concluded that the targets around the center of the screen are more reachable than the targets on the lower portion of the screen since there are higher throughput values for the center sequence of \textit{Alphabets} test (see Fig. \ref{fig:sensitivity_results_elapsed}). This is natural since pitching the head up or down is relatively more uncomfortable task than yawing the head left or right. Placing the frequently visited targets around the screen center while reserving the lower portion for information display or for less commonly used targets like the ones in our design (Fig. \ref{fig:Teaser}) would provide a more ergonomic interaction via head-based pointing on mobile and keeps the amount of uncomfortable pitching tasks to the minimum.

We highlight the potential of the proposed approach as an important number of people who experienced our approach considered it useful for them in several use cases such as baby feeding or fixing automobile with dirty hands. In addition to able-bodied user group, our single participant with motor impairments was also able to practice precise target selection on such a small screen for the very first time. He also reported that a hands-free interaction with smart phones is especially a serious need for any wheelchair rider even if they have fine hand control since they also would like to keep interacting with their phones while moving. In so many cases like this, our open-sourced hands-free UI components would be really helpful for developers to build new applications in consideration or these needs. Therefore, future work would be developing any kind of mobile applications on top of our framework or adding new head-pointing sensitive UI components to our open-sourced tool set. 
\section{Conclusion}
In this work, we proposed an open-source solution for mobile phones to enable an easy-to-use interface for buttons that are sensitive to head pointing. 

We hope that this open source will make it easier for developers to incorporate head tracking into  applications not limited to eCommerce. This includes socially impactful applications for users with motor control disabilities. It also allows hands-free applications such as following recipe while cooking or fixing car while following an instructional manual. Fun applications like gaming can also consume this.

\newpage

\bibliographystyle{ACM-Reference-Format}
\bibliography{references}


\begin{thebibliography}{66}


\ifx \showCODEN    \undefined \def \showCODEN     #1{\unskip}     \fi
\ifx \showDOI      \undefined \def \showDOI       #1{#1}\fi
\ifx \showISBNx    \undefined \def \showISBNx     #1{\unskip}     \fi
\ifx \showISBNxiii \undefined \def \showISBNxiii  #1{\unskip}     \fi
\ifx \showISSN     \undefined \def \showISSN      #1{\unskip}     \fi
\ifx \showLCCN     \undefined \def \showLCCN      #1{\unskip}     \fi
\ifx \shownote     \undefined \def \shownote      #1{#1}          \fi
\ifx \showarticletitle \undefined \def \showarticletitle #1{#1}   \fi
\ifx \showURL      \undefined \def \showURL       {\relax}        \fi
\providecommand\bibfield[2]{#2}
\providecommand\bibinfo[2]{#2}
\providecommand\natexlab[1]{#1}
\providecommand\showeprint[2][]{arXiv:#2}

\bibitem[\protect\citeauthoryear{Abbaszadegan, Yaghoubi, and
  MacKenzie}{Abbaszadegan et~al\mbox{.}}{2018}]%
        {abbaszadegan2018trackmaze}
\bibfield{author}{\bibinfo{person}{Mahdieh Abbaszadegan},
  \bibinfo{person}{Sohrab Yaghoubi}, {and} \bibinfo{person}{I~Scott
  MacKenzie}.} \bibinfo{year}{2018}\natexlab{}.
\newblock \showarticletitle{TrackMaze: A Comparison of Head-Tracking,
  Eye-Tracking, and Tilt as Input Methods for Mobile Games}. In
  \bibinfo{booktitle}{\emph{International Conference on Human-Computer
  Interaction}}. Springer, \bibinfo{pages}{393--405}.
\newblock
\urldef\tempurl%
\url{https://doi.org/10.1007/978-3-319-91250-9_31}
\showDOI{\tempurl}


\bibitem[\protect\citeauthoryear{Accessibility}{Accessibility}{2018}]%
        {essentialaccessibility}
\bibfield{author}{\bibinfo{person}{Essential Accessibility}.}
  \bibinfo{year}{2018}\natexlab{}.
\newblock \bibinfo{title}{Essential Accessibility}.
\newblock
\newblock
\newblock
\shownote{Retrieved August 6, 2018 from
  \url{https://www.essentialaccessibility.com/assistive-technology-for-android/}.}


\bibitem[\protect\citeauthoryear{Barfield}{Barfield}{2015}]%
        {barfield2015fundamentals}
\bibfield{author}{\bibinfo{person}{Woodrow Barfield}.}
  \bibinfo{year}{2015}\natexlab{}.
\newblock \bibinfo{booktitle}{\emph{Fundamentals of wearable computers and
  augmented reality}}.
\newblock \bibinfo{publisher}{CRC Press}.
\newblock


\bibitem[\protect\citeauthoryear{Bates and Istance}{Bates and Istance}{2003}]%
        {bates2003eye}
\bibfield{author}{\bibinfo{person}{Richard Bates} {and}
  \bibinfo{person}{Howell~O Istance}.} \bibinfo{year}{2003}\natexlab{}.
\newblock \showarticletitle{Why are eye mice unpopular? A detailed comparison
  of head and eye controlled assistive technology pointing devices}.
\newblock \bibinfo{journal}{\emph{Universal Access in the Information Society}}
  \bibinfo{volume}{2}, \bibinfo{number}{3} (\bibinfo{year}{2003}),
  \bibinfo{pages}{280--290}.
\newblock
\urldef\tempurl%
\url{https://doi.org/10.1007/s10209-003-0053-y}
\showDOI{\tempurl}


\bibitem[\protect\citeauthoryear{Betke, Gips, and Fleming}{Betke
  et~al\mbox{.}}{2002}]%
        {betke2002camera}
\bibfield{author}{\bibinfo{person}{Margrit Betke}, \bibinfo{person}{James
  Gips}, {and} \bibinfo{person}{Peter Fleming}.}
  \bibinfo{year}{2002}\natexlab{}.
\newblock \showarticletitle{The camera mouse: Visual tracking of body features
  to provide computer access for people with severe disabilities}.
\newblock \bibinfo{journal}{\emph{IEEE Transactions on neural systems and
  Rehabilitation Engineering}} \bibinfo{volume}{10}, \bibinfo{number}{1}
  (\bibinfo{year}{2002}), \bibinfo{pages}{1--10}.
\newblock
\urldef\tempurl%
\url{https://doi.org/10.1109/TNSRE.2002.1021581}
\showDOI{\tempurl}


\bibitem[\protect\citeauthoryear{Bichsel and Pentland}{Bichsel and
  Pentland}{[n. d.]}]%
        {bichsel1993automatic}
\bibfield{author}{\bibinfo{person}{Martin Bichsel} {and} \bibinfo{person}{Alex
  Pentland}.} \bibinfo{year}{[n. d.]}\natexlab{}.
\newblock \showarticletitle{Automatic interpretation of human head movements}.
\newblock


\bibitem[\protect\citeauthoryear{Billinghurst, Clark, Lee,
  et~al\mbox{.}}{Billinghurst et~al\mbox{.}}{2015}]%
        {billinghurst2015survey}
\bibfield{author}{\bibinfo{person}{Mark Billinghurst}, \bibinfo{person}{Adrian
  Clark}, \bibinfo{person}{Gun Lee}, {et~al\mbox{.}}}
  \bibinfo{year}{2015}\natexlab{}.
\newblock \showarticletitle{A survey of augmented reality}.
\newblock \bibinfo{journal}{\emph{Foundations and Trends{\textregistered} in
  Human--Computer Interaction}} \bibinfo{volume}{8}, \bibinfo{number}{2-3}
  (\bibinfo{year}{2015}), \bibinfo{pages}{73--272}.
\newblock
\urldef\tempurl%
\url{https://doi.org/10.1561/1100000049}
\showDOI{\tempurl}


\bibitem[\protect\citeauthoryear{Bowman, Kruijff, LaViola~Jr, and
  Poupyrev}{Bowman et~al\mbox{.}}{2004}]%
        {bowman20043d}
\bibfield{author}{\bibinfo{person}{Doug Bowman}, \bibinfo{person}{Ernst
  Kruijff}, \bibinfo{person}{Joseph~J LaViola~Jr}, {and}
  \bibinfo{person}{Ivan~P Poupyrev}.} \bibinfo{year}{2004}\natexlab{}.
\newblock \bibinfo{booktitle}{\emph{3D User interfaces: theory and practice}}.
\newblock \bibinfo{publisher}{Addison Wesley}.
\newblock
\urldef\tempurl%
\url{https://doi.org/10.1162/pres.2005.14.1.117}
\showDOI{\tempurl}


\bibitem[\protect\citeauthoryear{Brewster, Lumsden, Bell, Hall, and
  Tasker}{Brewster et~al\mbox{.}}{2003}]%
        {brewster2003multimodal}
\bibfield{author}{\bibinfo{person}{Stephen Brewster}, \bibinfo{person}{Joanna
  Lumsden}, \bibinfo{person}{Marek Bell}, \bibinfo{person}{Malcolm Hall}, {and}
  \bibinfo{person}{Stuart Tasker}.} \bibinfo{year}{2003}\natexlab{}.
\newblock \showarticletitle{Multimodal 'eyes-free' interaction techniques for
  wearable devices}. In \bibinfo{booktitle}{\emph{Proceedings of the SIGCHI
  conference on Human factors in computing systems}}. ACM,
  \bibinfo{pages}{473--480}.
\newblock
\urldef\tempurl%
\url{https://doi.org/10.1145/642693.642694}
\showDOI{\tempurl}


\bibitem[\protect\citeauthoryear{Canessa, Chessa, Gibaldi, Sabatini, and
  Solari}{Canessa et~al\mbox{.}}{2014}]%
        {canessa2014calibrated}
\bibfield{author}{\bibinfo{person}{Andrea Canessa}, \bibinfo{person}{Manuela
  Chessa}, \bibinfo{person}{Agostino Gibaldi}, \bibinfo{person}{Silvio~P
  Sabatini}, {and} \bibinfo{person}{Fabio Solari}.}
  \bibinfo{year}{2014}\natexlab{}.
\newblock \showarticletitle{Calibrated depth and color cameras for accurate 3D
  interaction in a stereoscopic augmented reality environment}.
\newblock \bibinfo{journal}{\emph{Journal of Visual Communication and Image
  Representation}} \bibinfo{volume}{25}, \bibinfo{number}{1}
  (\bibinfo{year}{2014}), \bibinfo{pages}{227--237}.
\newblock
\urldef\tempurl%
\url{https://doi.org/10.1016/j.jvcir.2013.02.011}
\showDOI{\tempurl}


\bibitem[\protect\citeauthoryear{Chen, Lee, and Lin}{Chen
  et~al\mbox{.}}{2015}]%
        {chen2015augmented}
\bibfield{author}{\bibinfo{person}{Chien-Hsu Chen}, \bibinfo{person}{I-Jui
  Lee}, {and} \bibinfo{person}{Ling-Yi Lin}.} \bibinfo{year}{2015}\natexlab{}.
\newblock \showarticletitle{Augmented reality-based self-facial modeling to
  promote the emotional expression and social skills of adolescents with autism
  spectrum disorders}.
\newblock \bibinfo{journal}{\emph{Research in developmental disabilities}}
  \bibinfo{volume}{36} (\bibinfo{year}{2015}), \bibinfo{pages}{396--403}.
\newblock
\urldef\tempurl%
\url{https://doi.org/10.1016/j.ridd.2014.10.015}
\showDOI{\tempurl}


\bibitem[\protect\citeauthoryear{Clifford, Tuanquin, and Lindeman}{Clifford
  et~al\mbox{.}}{2017}]%
        {clifford2017jedi}
\bibfield{author}{\bibinfo{person}{Rory~MS Clifford}, \bibinfo{person}{Nikita
  Mae~B Tuanquin}, {and} \bibinfo{person}{Robert~W Lindeman}.}
  \bibinfo{year}{2017}\natexlab{}.
\newblock \showarticletitle{Jedi ForceExtension: Telekinesis as a Virtual
  Reality interaction metaphor}. In \bibinfo{booktitle}{\emph{3D User
  Interfaces (3DUI), 2017 IEEE Symposium on}}. IEEE, \bibinfo{pages}{239--240}.
\newblock
\urldef\tempurl%
\url{https://doi.org/10.1109/3DUI.2017.7893360}
\showDOI{\tempurl}


\bibitem[\protect\citeauthoryear{Corporation}{Corporation}{2017}]%
        {headmouse}
\bibfield{author}{\bibinfo{person}{Origin~Instruments Corporation}.}
  \bibinfo{year}{2017}\natexlab{}.
\newblock \bibinfo{title}{HeadMouse Nano}.
\newblock
\newblock
\newblock
\shownote{Retrieved July 17, 2018 from
  \url{http://www.orin.com/access/headmouse/}.}


\bibitem[\protect\citeauthoryear{Dynavox}{Dynavox}{2018}]%
        {tobiidynavox}
\bibfield{author}{\bibinfo{person}{Tobii Dynavox}.}
  \bibinfo{year}{2018}\natexlab{}.
\newblock \bibinfo{title}{Microsoft \& Tobii Dynavox}.
\newblock
\newblock
\newblock
\shownote{Retrieved July 16, 2018 from
  \url{https://www.tobiidynavox.com/en-US/landing-pages/td_and_microsoft/}.}


\bibitem[\protect\citeauthoryear{Findlater, Moffatt, Froehlich, Malu, and
  Zhang}{Findlater et~al\mbox{.}}{2017}]%
        {findlater2017comparing}
\bibfield{author}{\bibinfo{person}{Leah Findlater}, \bibinfo{person}{Karyn
  Moffatt}, \bibinfo{person}{Jon~E Froehlich}, \bibinfo{person}{Meethu Malu},
  {and} \bibinfo{person}{Joan Zhang}.} \bibinfo{year}{2017}\natexlab{}.
\newblock \showarticletitle{Comparing touchscreen and mouse input performance
  by people with and without upper body motor impairments}. In
  \bibinfo{booktitle}{\emph{Proceedings of the 2017 CHI Conference on Human
  Factors in Computing Systems}}. ACM, \bibinfo{pages}{6056--6061}.
\newblock
\urldef\tempurl%
\url{https://doi.org/10.1145/3025453.3025603}
\showDOI{\tempurl}


\bibitem[\protect\citeauthoryear{Fitts}{Fitts}{1954}]%
        {fitts1954information}
\bibfield{author}{\bibinfo{person}{Paul~M Fitts}.}
  \bibinfo{year}{1954}\natexlab{}.
\newblock \showarticletitle{The information capacity of the human motor system
  in controlling the amplitude of movement}.
\newblock \bibinfo{journal}{\emph{Journal of experimental psychology}}
  \bibinfo{volume}{47}, \bibinfo{number}{6} (\bibinfo{year}{1954}),
  \bibinfo{pages}{381}.
\newblock
\urldef\tempurl%
\url{https://doi.org/10.1037/h0055392}
\showDOI{\tempurl}


\bibitem[\protect\citeauthoryear{for Health~Statistics}{for
  Health~Statistics}{2017}]%
        {disabilities}
\bibfield{author}{\bibinfo{person}{National~Center for Health~Statistics}.}
  \bibinfo{year}{2017}\natexlab{}.
\newblock \bibinfo{title}{Disability and Functioning (Noninstitutionalized
  Adults Aged 18 and Over)}.
\newblock
\newblock
\newblock
\shownote{Retrieved July 9, 2018 from
  \url{https://www.cdc.gov/nchs/fastats/disability.htm}.}


\bibitem[\protect\citeauthoryear{Gizatdinova, {\v{S}}pakov, and
  Surakka}{Gizatdinova et~al\mbox{.}}{2012}]%
        {gizatdinova2012comparison}
\bibfield{author}{\bibinfo{person}{Yulia Gizatdinova}, \bibinfo{person}{Oleg
  {\v{S}}pakov}, {and} \bibinfo{person}{Veikko Surakka}.}
  \bibinfo{year}{2012}\natexlab{}.
\newblock \showarticletitle{Comparison of video-based pointing and selection
  techniques for hands-free text entry}. In
  \bibinfo{booktitle}{\emph{Proceedings of the international working conference
  on advanced visual interfaces}}. ACM, \bibinfo{pages}{132--139}.
\newblock
\urldef\tempurl%
\url{https://doi.org/10.1145/2254556.2254582}
\showDOI{\tempurl}


\bibitem[\protect\citeauthoryear{Glassouse}{Glassouse}{2018}]%
        {glassouse}
\bibfield{author}{\bibinfo{person}{Glassouse}.}
  \bibinfo{year}{2018}\natexlab{}.
\newblock \bibinfo{title}{Glassouse Assistive Device}.
\newblock
\newblock
\newblock
\shownote{Retrieved July 17, 2018 from \url{http://glassouse.com/}.}


\bibitem[\protect\citeauthoryear{Huang, Veeraraghavan, and Sabharwal}{Huang
  et~al\mbox{.}}{2015}]%
        {huang2015tabletgaze}
\bibfield{author}{\bibinfo{person}{Qiong Huang}, \bibinfo{person}{Ashok
  Veeraraghavan}, {and} \bibinfo{person}{Ashutosh Sabharwal}.}
  \bibinfo{year}{2015}\natexlab{}.
\newblock \showarticletitle{TabletGaze: unconstrained appearance-based gaze
  estimation in mobile tablets}.
\newblock \bibinfo{journal}{\emph{arXiv preprint arXiv:1508.01244}}
  (\bibinfo{year}{2015}).
\newblock
\urldef\tempurl%
\url{https://doi.org/10.1007/s00138-017-0852-4}
\showDOI{\tempurl}


\bibitem[\protect\citeauthoryear{Inc.}{Inc.}{2018a}]%
        {arkit}
\bibfield{author}{\bibinfo{person}{Apple Inc.}}
  \bibinfo{year}{2018}\natexlab{a}.
\newblock \bibinfo{title}{Get Ready for ARKit 2}.
\newblock
\newblock
\newblock
\shownote{Retrieved July 17, 2018 from
  \url{https://developer.apple.com/arkit/}.}


\bibitem[\protect\citeauthoryear{Inc.}{Inc.}{2018b}]%
        {truedepth}
\bibfield{author}{\bibinfo{person}{Apple Inc.}}
  \bibinfo{year}{2018}\natexlab{b}.
\newblock \bibinfo{title}{TrueDepth Camera}.
\newblock
\newblock
\newblock
\shownote{Retrieved July 17, 2018 from
  \url{https://www.apple.com/iphone-x/\#truedepth-camera}.}


\bibitem[\protect\citeauthoryear{Inc.}{Inc.}{2018c}]%
        {switchcontrol}
\bibfield{author}{\bibinfo{person}{Apple Inc.}}
  \bibinfo{year}{2018}\natexlab{c}.
\newblock \bibinfo{title}{Use Switch Control to navigate your iPhone, iPad, or
  iPod touch}.
\newblock
\newblock
\newblock
\shownote{Retrieved July 15, 2018 from
  \url{https://support.apple.com/en-us/ht201370}.}


\bibitem[\protect\citeauthoryear{ISO}{ISO}{[n. d.]}]%
        {iso9241}
\bibfield{author}{\bibinfo{person}{ISO ISO}.} \bibinfo{year}{[n.
  d.]}\natexlab{}.
\newblock \showarticletitle{9241-9 Ergonomic requirements for office work with
  visual display terminals (VDTs)-Part 9: Requirements for non-keyboard input
  devices (FDIS-Final Draft International Standard), 2000}.
\newblock \bibinfo{journal}{\emph{International Organization for
  Standardization}} (\bibinfo{year}{[n. d.]}).
\newblock


\bibitem[\protect\citeauthoryear{{KPMG}}{{KPMG}}{2017}]%
        {kpmg2017truth}
\bibfield{author}{\bibinfo{person}{{KPMG}}.} \bibinfo{year}{2017}\natexlab{}.
\newblock \bibinfo{booktitle}{\emph{The Truth About Online Consumers}}.
\newblock \bibinfo{type}{2017 Global Online Consumer Report}.
  \bibinfo{institution}{Klynveld Peat Marwick Goerdeler}.
\newblock


\bibitem[\protect\citeauthoryear{Krafka, Khosla, Kellnhofer, Kannan,
  Bhandarkar, Matusik, and Torralba}{Krafka et~al\mbox{.}}{2016}]%
        {krafka2016eye}
\bibfield{author}{\bibinfo{person}{Kyle Krafka}, \bibinfo{person}{Aditya
  Khosla}, \bibinfo{person}{Petr Kellnhofer}, \bibinfo{person}{Harini Kannan},
  \bibinfo{person}{Suchendra Bhandarkar}, \bibinfo{person}{Wojciech Matusik},
  {and} \bibinfo{person}{Antonio Torralba}.} \bibinfo{year}{2016}\natexlab{}.
\newblock \showarticletitle{Eye tracking for everyone}. In
  \bibinfo{booktitle}{\emph{Proceedings of the IEEE conference on computer
  vision and pattern recognition}}. \bibinfo{pages}{2176--2184}.
\newblock
\urldef\tempurl%
\url{https://doi.org/10.1109/CVPR.2016.239}
\showDOI{\tempurl}


\bibitem[\protect\citeauthoryear{Kurauchi, Feng, Joshi, Morimoto, and
  Betke}{Kurauchi et~al\mbox{.}}{2016}]%
        {kurauchi2016eyeswipe}
\bibfield{author}{\bibinfo{person}{Andrew Kurauchi}, \bibinfo{person}{Wenxin
  Feng}, \bibinfo{person}{Ajjen Joshi}, \bibinfo{person}{Carlos Morimoto},
  {and} \bibinfo{person}{Margrit Betke}.} \bibinfo{year}{2016}\natexlab{}.
\newblock \showarticletitle{EyeSwipe: Dwell-free text entry using gaze paths}.
  In \bibinfo{booktitle}{\emph{Proceedings of the 2016 CHI Conference on Human
  Factors in Computing Systems}}. ACM, \bibinfo{pages}{1952--1956}.
\newblock
\urldef\tempurl%
\url{https://doi.org/10.1145/2858036.2858335}
\showDOI{\tempurl}


\bibitem[\protect\citeauthoryear{Kyt{\"o}, Ens, Piumsomboon, Lee, and
  Billinghurst}{Kyt{\"o} et~al\mbox{.}}{2018}]%
        {kyto2018pinpointing}
\bibfield{author}{\bibinfo{person}{Mikko Kyt{\"o}}, \bibinfo{person}{Barrett
  Ens}, \bibinfo{person}{Thammathip Piumsomboon}, \bibinfo{person}{Gun~A Lee},
  {and} \bibinfo{person}{Mark Billinghurst}.} \bibinfo{year}{2018}\natexlab{}.
\newblock \showarticletitle{Pinpointing: Precise Head-and Eye-Based Target
  Selection for Augmented Reality}. In \bibinfo{booktitle}{\emph{Proceedings of
  the 2018 CHI Conference on Human Factors in Computing Systems}}. ACM,
  \bibinfo{pages}{81}.
\newblock
\urldef\tempurl%
\url{https://doi.org/10.1145/3173574.3173655}
\showDOI{\tempurl}


\bibitem[\protect\citeauthoryear{LLC}{LLC}{2018}]%
        {arcore}
\bibfield{author}{\bibinfo{person}{Google LLC}.}
  \bibinfo{year}{2018}\natexlab{}.
\newblock \bibinfo{title}{ARCore Overview}.
\newblock
\newblock
\newblock
\shownote{Retrieved July 17, 2018 from
  \url{https://developers.google.com/ar/discover/}.}


\bibitem[\protect\citeauthoryear{LLC}{LLC}{2016}]%
        {smylemouse}
\bibfield{author}{\bibinfo{person}{Perceptive~Devices LLC}.}
  \bibinfo{year}{2016}\natexlab{}.
\newblock \bibinfo{title}{SmyleMouse}.
\newblock
\newblock
\newblock
\shownote{Retrieved July 15, 2018 from \url{https://smylemouse.com/}.}


\bibitem[\protect\citeauthoryear{Lv, Halawani, Feng, Ur~R{\'e}hman, and Li}{Lv
  et~al\mbox{.}}{2015}]%
        {lv2015touch}
\bibfield{author}{\bibinfo{person}{Zhihan Lv}, \bibinfo{person}{Alaa Halawani},
  \bibinfo{person}{Shengzhong Feng}, \bibinfo{person}{Shafiq Ur~R{\'e}hman},
  {and} \bibinfo{person}{Haibo Li}.} \bibinfo{year}{2015}\natexlab{}.
\newblock \showarticletitle{Touch-less interactive augmented reality game on
  vision-based wearable device}.
\newblock \bibinfo{journal}{\emph{Personal and Ubiquitous Computing}}
  \bibinfo{volume}{19}, \bibinfo{number}{3-4} (\bibinfo{year}{2015}),
  \bibinfo{pages}{551--567}.
\newblock
\urldef\tempurl%
\url{https://doi.org/10.1007/s00779-015-0844-1}
\showDOI{\tempurl}


\bibitem[\protect\citeauthoryear{MacKenzie}{MacKenzie}{2018}]%
        {mackenzie2018fitts}
\bibfield{author}{\bibinfo{person}{I~Scott MacKenzie}.}
  \bibinfo{year}{2018}\natexlab{}.
\newblock \showarticletitle{Fitts’ Law}.
\newblock \bibinfo{journal}{\emph{The Wiley Handbook of Human Computer
  Interaction}}  \bibinfo{volume}{1} (\bibinfo{year}{2018}),
  \bibinfo{pages}{347--370}.
\newblock
\urldef\tempurl%
\url{https://doi.org/10.1002/9781118976005.ch17}
\showDOI{\tempurl}


\bibitem[\protect\citeauthoryear{Magee, Felzer, and MacKenzie}{Magee
  et~al\mbox{.}}{2015}]%
        {magee2015camera}
\bibfield{author}{\bibinfo{person}{John Magee}, \bibinfo{person}{Torsten
  Felzer}, {and} \bibinfo{person}{I~Scott MacKenzie}.}
  \bibinfo{year}{2015}\natexlab{}.
\newblock \showarticletitle{Camera Mouse+ ClickerAID: Dwell vs. single-muscle
  click actuation in mouse-replacement interfaces}. In
  \bibinfo{booktitle}{\emph{International Conference on Universal Access in
  Human-Computer Interaction}}. Springer, \bibinfo{pages}{74--84}.
\newblock
\urldef\tempurl%
\url{https://doi.org/10.1007/978-3-319-20678-3_8}
\showDOI{\tempurl}


\bibitem[\protect\citeauthoryear{Majaranta}{Majaranta}{2011}]%
        {majaranta2011gaze}
\bibfield{author}{\bibinfo{person}{P{\"a}ivi Majaranta}.}
  \bibinfo{year}{2011}\natexlab{}.
\newblock \bibinfo{booktitle}{\emph{Gaze Interaction and Applications of Eye
  Tracking: Advances in Assistive Technologies: Advances in Assistive
  Technologies}}.
\newblock \bibinfo{publisher}{IGI Global}.
\newblock
\urldef\tempurl%
\url{https://doi.org/10.4018/978-1-61350-098-9}
\showDOI{\tempurl}


\bibitem[\protect\citeauthoryear{Manresa-Yee, Ponsa, Varona, and
  Perales}{Manresa-Yee et~al\mbox{.}}{2010}]%
        {manresa2010user}
\bibfield{author}{\bibinfo{person}{Cristina Manresa-Yee}, \bibinfo{person}{Pere
  Ponsa}, \bibinfo{person}{Javier Varona}, {and} \bibinfo{person}{Francisco~J
  Perales}.} \bibinfo{year}{2010}\natexlab{}.
\newblock \showarticletitle{User experience to improve the usability of a
  vision-based interface}.
\newblock \bibinfo{journal}{\emph{Interacting with Computers}}
  \bibinfo{volume}{22}, \bibinfo{number}{6} (\bibinfo{year}{2010}),
  \bibinfo{pages}{594--605}.
\newblock
\urldef\tempurl%
\url{https://doi.org/10.1016/j.intcom.2010.06.004}
\showDOI{\tempurl}


\bibitem[\protect\citeauthoryear{Mauri}{Mauri}{2017}]%
        {viacam}
\bibfield{author}{\bibinfo{person}{Cesar Mauri}.}
  \bibinfo{year}{2017}\natexlab{}.
\newblock \bibinfo{title}{Enable Viacam}.
\newblock
\newblock
\newblock
\shownote{Retrieved July 15, 2018 from
  \url{http://eviacam.crea-si.com/index.php}.}


\bibitem[\protect\citeauthoryear{Mauri}{Mauri}{2018}]%
        {evafacialmouse}
\bibfield{author}{\bibinfo{person}{Cesar Mauri}.}
  \bibinfo{year}{2018}\natexlab{}.
\newblock \bibinfo{title}{EVA Facial Mouse}.
\newblock
\newblock
\newblock
\shownote{Retrieved July 16, 2018 from
  \url{https://github.com/cmauri/eva_facial_mouse\#user-content-eva-facial-mouse}.}


\bibitem[\protect\citeauthoryear{Mauri, Granollers~i Saltiveri,
  Lor{\'e}s~Vidal, and Garc{\'\i}a}{Mauri et~al\mbox{.}}{2006}]%
        {mauri2006computer}
\bibfield{author}{\bibinfo{person}{C{\'e}sar Mauri}, \bibinfo{person}{Toni
  Granollers~i Saltiveri}, \bibinfo{person}{Jes{\'u}s Lor{\'e}s~Vidal}, {and}
  \bibinfo{person}{Mabel Garc{\'\i}a}.} \bibinfo{year}{2006}\natexlab{}.
\newblock \showarticletitle{Computer vision interaction for people with severe
  movement restrictions}.
\newblock \bibinfo{journal}{\emph{Human Technology: An Interdisciplinary
  Journal on Humans in ICT Environments, vol. 2, n{\'u}m. 1, p. 38-54}}
  (\bibinfo{year}{2006}).
\newblock
\urldef\tempurl%
\url{https://doi.org/10.17011/ht/urn.2006158}
\showDOI{\tempurl}


\bibitem[\protect\citeauthoryear{Microsoft}{Microsoft}{2018}]%
        {eyecontrol}
\bibfield{author}{\bibinfo{person}{Microsoft}.}
  \bibinfo{year}{2018}\natexlab{}.
\newblock \bibinfo{title}{Eye Control for Windows 10}.
\newblock
\newblock
\newblock
\shownote{Retrieved July 16, 2018 from
  \url{https://www.microsoft.com/en-us/garage/wall-of-fame/eye-control-windows-10/}.}


\bibitem[\protect\citeauthoryear{Montague, Nicolau, and Hanson}{Montague
  et~al\mbox{.}}{2014}]%
        {montague2014motor}
\bibfield{author}{\bibinfo{person}{Kyle Montague}, \bibinfo{person}{Hugo
  Nicolau}, {and} \bibinfo{person}{Vicki~L Hanson}.}
  \bibinfo{year}{2014}\natexlab{}.
\newblock \showarticletitle{Motor-impaired touchscreen interactions in the
  wild}. In \bibinfo{booktitle}{\emph{Proceedings of the 16th international ACM
  SIGACCESS conference on Computers \& accessibility}}. ACM,
  \bibinfo{pages}{123--130}.
\newblock
\urldef\tempurl%
\url{https://doi.org/10.1145/2661334.2661362}
\showDOI{\tempurl}


\bibitem[\protect\citeauthoryear{Mott, Vatavu, Kane, and Wobbrock}{Mott
  et~al\mbox{.}}{2016}]%
        {mott2016smart}
\bibfield{author}{\bibinfo{person}{Martez~E Mott}, \bibinfo{person}{Radu-Daniel
  Vatavu}, \bibinfo{person}{Shaun~K Kane}, {and} \bibinfo{person}{Jacob~O
  Wobbrock}.} \bibinfo{year}{2016}\natexlab{}.
\newblock \showarticletitle{Smart touch: Improving touch accuracy for people
  with motor impairments with template matching}. In
  \bibinfo{booktitle}{\emph{Proceedings of the 2016 CHI Conference on Human
  Factors in Computing Systems}}. ACM, \bibinfo{pages}{1934--1946}.
\newblock
\urldef\tempurl%
\url{https://doi.org/10.1145/2858036.2858390}
\showDOI{\tempurl}


\bibitem[\protect\citeauthoryear{MyGaze}{MyGaze}{2018}]%
        {mygaze}
\bibfield{author}{\bibinfo{person}{MyGaze}.} \bibinfo{year}{2018}\natexlab{}.
\newblock \bibinfo{title}{MyGaze Assistive}.
\newblock
\newblock
\newblock
\shownote{Retrieved July 16, 2018 from
  \url{http://www.mygaze.com/products/mygaze-assistive/}.}


\bibitem[\protect\citeauthoryear{{Neil Mawston}}{{Neil Mawston}}{2018}]%
        {StrategyAnalytics2018iPhoneX}
\bibfield{author}{\bibinfo{person}{{Neil Mawston}}.}
  \bibinfo{year}{2018}\natexlab{}.
\newblock \bibinfo{booktitle}{\emph{Apple iPhone X Becomes World’s No.1
  Smartphone Model in Q1 2018}}.
\newblock \bibinfo{type}{{T}echnical {R}eport}. \bibinfo{institution}{Strategy
  Analytics}.
\newblock
\newblock
\shownote{Retrieved August 13, 2018 from
  \url{https://www.strategyanalytics.com/access-services/devices/mobile-phones/smartphone/smartphone-model-tracker/reports/report-detail/apple-iphone-x-becomes-world-s-best-selling-smartphone-model-in-q1-2018}.}


\bibitem[\protect\citeauthoryear{of~Boston~College}{of~Boston~College}{2018}]%
        {cameramouse}
\bibfield{author}{\bibinfo{person}{Trustees of Boston~College}.}
  \bibinfo{year}{2018}\natexlab{}.
\newblock \bibinfo{title}{CameraMouse}.
\newblock
\newblock
\newblock
\shownote{Retrieved July 15, 2018 from \url{http://www.cameramouse.org/}.}


\bibitem[\protect\citeauthoryear{oy.}{oy.}{2018}]%
        {quhazono}
\bibfield{author}{\bibinfo{person}{Quha oy.}} \bibinfo{year}{2018}\natexlab{}.
\newblock \bibinfo{title}{Quha Zono}.
\newblock
\newblock
\newblock
\shownote{Retrieved July 15, 2018 from
  \url{http://www.quha.com/products-2/zono/}.}


\bibitem[\protect\citeauthoryear{Penkar, Lutteroth, and Weber}{Penkar
  et~al\mbox{.}}{2012}]%
        {penkar2012designing}
\bibfield{author}{\bibinfo{person}{Abdul~Moiz Penkar},
  \bibinfo{person}{Christof Lutteroth}, {and} \bibinfo{person}{Gerald Weber}.}
  \bibinfo{year}{2012}\natexlab{}.
\newblock \showarticletitle{Designing for the eye: design parameters for dwell
  in gaze interaction}. In \bibinfo{booktitle}{\emph{Proceedings of the 24th
  Australian Computer-Human Interaction Conference}}. ACM,
  \bibinfo{pages}{479--488}.
\newblock
\urldef\tempurl%
\url{https://doi.org/10.1145/2414536.2414609}
\showDOI{\tempurl}


\bibitem[\protect\citeauthoryear{Perea~y Monsuw{\'e}, Dellaert, and
  De~Ruyter}{Perea~y Monsuw{\'e} et~al\mbox{.}}{2004}]%
        {perea2004drives}
\bibfield{author}{\bibinfo{person}{To{\~n}ita Perea~y Monsuw{\'e}},
  \bibinfo{person}{Benedict~GC Dellaert}, {and} \bibinfo{person}{Ko
  De~Ruyter}.} \bibinfo{year}{2004}\natexlab{}.
\newblock \showarticletitle{What drives consumers to shop online? A literature
  review}.
\newblock \bibinfo{journal}{\emph{International journal of service industry
  management}} \bibinfo{volume}{15}, \bibinfo{number}{1}
  (\bibinfo{year}{2004}), \bibinfo{pages}{102--121}.
\newblock
\urldef\tempurl%
\url{https://doi.org/10.1108/09564230410523358}
\showDOI{\tempurl}


\bibitem[\protect\citeauthoryear{Polacek, Grill, and Tscheligi}{Polacek
  et~al\mbox{.}}{2013}]%
        {polacek2013nosetapping}
\bibfield{author}{\bibinfo{person}{Ondrej Polacek}, \bibinfo{person}{Thomas
  Grill}, {and} \bibinfo{person}{Manfred Tscheligi}.}
  \bibinfo{year}{2013}\natexlab{}.
\newblock \showarticletitle{NoseTapping: what else can you do with your nose?}.
  In \bibinfo{booktitle}{\emph{Proceedings of the 12th International Conference
  on Mobile and Ubiquitous Multimedia}}. ACM, \bibinfo{pages}{32}.
\newblock
\urldef\tempurl%
\url{https://doi.org/10.1145/2541831.2541867}
\showDOI{\tempurl}


\bibitem[\protect\citeauthoryear{Ranjan, De~Mello, and Kautz}{Ranjan
  et~al\mbox{.}}{2018}]%
        {ranjan2018light}
\bibfield{author}{\bibinfo{person}{Rajeev Ranjan}, \bibinfo{person}{Shalini
  De~Mello}, {and} \bibinfo{person}{Jan Kautz}.}
  \bibinfo{year}{2018}\natexlab{}.
\newblock \showarticletitle{Light-weight Head Pose Invariant Gaze Tracking}.
\newblock \bibinfo{journal}{\emph{arXiv preprint arXiv:1804.08572}}
  (\bibinfo{year}{2018}).
\newblock


\bibitem[\protect\citeauthoryear{Riviere and Thakor}{Riviere and
  Thakor}{1996}]%
        {riviere1996effects}
\bibfield{author}{\bibinfo{person}{Cameron~N Riviere} {and}
  \bibinfo{person}{Nitish~V Thakor}.} \bibinfo{year}{1996}\natexlab{}.
\newblock \showarticletitle{Effects of age and disability on tracking tasks
  with a computer mouse: Accuracy and linearity}.
\newblock  (\bibinfo{year}{1996}).
\newblock
\newblock
\shownote{PMID: 8868412.}


\bibitem[\protect\citeauthoryear{Roig-Maim{\'o}, MacKenzie, Manresa-Yee, and
  Varona}{Roig-Maim{\'o} et~al\mbox{.}}{2017}]%
        {roig2017evaluating}
\bibfield{author}{\bibinfo{person}{Maria~Francesca Roig-Maim{\'o}},
  \bibinfo{person}{I~Scott MacKenzie}, \bibinfo{person}{Cristina Manresa-Yee},
  {and} \bibinfo{person}{Javier Varona}.} \bibinfo{year}{2017}\natexlab{}.
\newblock \showarticletitle{Evaluating fitts' law performance with a non-ISO
  task}. In \bibinfo{booktitle}{\emph{Proceedings of the XVIII International
  Conference on Human Computer Interaction}}. ACM, \bibinfo{pages}{5}.
\newblock
\urldef\tempurl%
\url{https://doi.org/10.1145/3123818.3123827}
\showDOI{\tempurl}


\bibitem[\protect\citeauthoryear{Roig-Maim{\'o}, MacKenzie, Manresa-Yee, and
  Varona}{Roig-Maim{\'o} et~al\mbox{.}}{2018}]%
        {roig2018head}
\bibfield{author}{\bibinfo{person}{Maria~Francesca Roig-Maim{\'o}},
  \bibinfo{person}{I~Scott MacKenzie}, \bibinfo{person}{Cristina Manresa-Yee},
  {and} \bibinfo{person}{Javier Varona}.} \bibinfo{year}{2018}\natexlab{}.
\newblock \showarticletitle{Head-tracking interfaces on mobile devices:
  Evaluation using Fitts’ law and a new multi-directional corner task for
  small displays}.
\newblock \bibinfo{journal}{\emph{International Journal of Human-Computer
  Studies}}  \bibinfo{volume}{112} (\bibinfo{year}{2018}),
  \bibinfo{pages}{1--15}.
\newblock
\urldef\tempurl%
\url{https://doi.org/10.1016/j.ijhcs.2017.12.003}
\showDOI{\tempurl}


\bibitem[\protect\citeauthoryear{Roig-Maim{\'o}, Manresa-Yee, Varona, and
  MacKenzie}{Roig-Maim{\'o} et~al\mbox{.}}{2016}]%
        {roig2016evaluation}
\bibfield{author}{\bibinfo{person}{Maria~Francesca Roig-Maim{\'o}},
  \bibinfo{person}{Cristina Manresa-Yee}, \bibinfo{person}{Javier Varona},
  {and} \bibinfo{person}{I~Scott MacKenzie}.} \bibinfo{year}{2016}\natexlab{}.
\newblock \showarticletitle{Evaluation of a mobile head-tracker interface for
  accessibility}. In \bibinfo{booktitle}{\emph{International Conference on
  Computers Helping People with Special Needs}}. Springer,
  \bibinfo{pages}{449--456}.
\newblock
\urldef\tempurl%
\url{https://doi.org/10.1007/978-3-319-41267-2_63}
\showDOI{\tempurl}


\bibitem[\protect\citeauthoryear{Smith and Anderson}{Smith and
  Anderson}{2016}]%
        {smith2016online}
\bibfield{author}{\bibinfo{person}{Aaron Smith} {and} \bibinfo{person}{Monica
  Anderson}.} \bibinfo{year}{2016}\natexlab{}.
\newblock \bibinfo{booktitle}{\emph{Online shopping and purchasing
  preferences}}.
\newblock \bibinfo{type}{{T}echnical {R}eport}. \bibinfo{institution}{Pew
  Research Center}.
\newblock
\urldef\tempurl%
\url{http://www.pewinternet.org/2016/12/19/online-shopping-and-purchasing-preferences/}
\showURL{%
\tempurl}


\bibitem[\protect\citeauthoryear{Speicher, Hell, Daiber, Simeone, and
  Kr{\"u}ger}{Speicher et~al\mbox{.}}{2018}]%
        {speicher2018virtual}
\bibfield{author}{\bibinfo{person}{Marco Speicher}, \bibinfo{person}{Philip
  Hell}, \bibinfo{person}{Florian Daiber}, \bibinfo{person}{Adalberto Simeone},
  {and} \bibinfo{person}{Antonio Kr{\"u}ger}.} \bibinfo{year}{2018}\natexlab{}.
\newblock \showarticletitle{A virtual reality shopping experience using the
  apartment metaphor}. In \bibinfo{booktitle}{\emph{Proceedings of the 2018
  International Conference on Advanced Visual Interfaces}}. ACM,
  \bibinfo{pages}{17}.
\newblock
\urldef\tempurl%
\url{https://doi.org/10.1145/3206505.3206518}
\showDOI{\tempurl}


\bibitem[\protect\citeauthoryear{Swaine, Labb{\'e}, Poldma, Barile, Fichten,
  Havel, Kehayia, Mazer, McKinley, and Rochette}{Swaine et~al\mbox{.}}{2014}]%
        {swaine2014exploring}
\bibfield{author}{\bibinfo{person}{Bonnie Swaine}, \bibinfo{person}{Delphine
  Labb{\'e}}, \bibinfo{person}{Tiiu Poldma}, \bibinfo{person}{Maria Barile},
  \bibinfo{person}{Catherine Fichten}, \bibinfo{person}{Alice Havel},
  \bibinfo{person}{Eva Kehayia}, \bibinfo{person}{Barbara Mazer},
  \bibinfo{person}{Patricia McKinley}, {and} \bibinfo{person}{Annie Rochette}.}
  \bibinfo{year}{2014}\natexlab{}.
\newblock \showarticletitle{Exploring the facilitators and barriers to shopping
  mall use by persons with disabilities and strategies for improvements:
  Perspectives from persons with disabilities, rehabilitation professionals and
  shopkeepers}.
\newblock \bibinfo{journal}{\emph{ALTER-European Journal of Disability
  Research/Revue Europ{\'e}enne de Recherche sur le Handicap}}
  \bibinfo{volume}{8}, \bibinfo{number}{3} (\bibinfo{year}{2014}),
  \bibinfo{pages}{217--229}.
\newblock
\urldef\tempurl%
\url{https://doi.org/10.1016/j.alter.2014.04.003}
\showDOI{\tempurl}


\bibitem[\protect\citeauthoryear{Tecla}{Tecla}{2018}]%
        {teclaswitch}
\bibfield{author}{\bibinfo{person}{Tecla}.} \bibinfo{year}{2018}\natexlab{}.
\newblock \bibinfo{title}{Introduction to 7 Common Adaptive Switches}.
\newblock
\newblock
\newblock
\shownote{Retrieved July 16, 2018 from
  \url{https://gettecla.com/blogs/news/introduction-to-assistive-switches}.}


\bibitem[\protect\citeauthoryear{{the Census Bureau}}{{the Census
  Bureau}}{2018}]%
        {Census2017quarterly}
\bibfield{author}{\bibinfo{person}{{the Census Bureau}}.}
  \bibinfo{year}{2018}\natexlab{}.
\newblock \bibinfo{booktitle}{\emph{Quarterly Retail E-Commerce Sales 1st
  Quarter 2018}}.
\newblock \bibinfo{type}{{T}echnical {R}eport}. \bibinfo{institution}{U.S.
  Department of Commerce}.
\newblock
\urldef\tempurl%
\url{https://www2.census.gov/retail/releases/historical/ecomm/18q1.pdf}
\showURL{%
\tempurl}


\bibitem[\protect\citeauthoryear{Thomas}{Thomas}{2012}]%
        {thomas2012survey}
\bibfield{author}{\bibinfo{person}{Bruce~H Thomas}.}
  \bibinfo{year}{2012}\natexlab{}.
\newblock \showarticletitle{A survey of visual, mixed, and augmented reality
  gaming}.
\newblock \bibinfo{journal}{\emph{Computers in Entertainment (CIE)}}
  \bibinfo{volume}{10}, \bibinfo{number}{1} (\bibinfo{year}{2012}),
  \bibinfo{pages}{3}.
\newblock
\urldef\tempurl%
\url{https://doi.org/10.1145/2381876.2381879}
\showDOI{\tempurl}


\bibitem[\protect\citeauthoryear{Turban, Outland, King, Lee, Liang, and
  Turban}{Turban et~al\mbox{.}}{2017}]%
        {turban2017electronic}
\bibfield{author}{\bibinfo{person}{Efraim Turban}, \bibinfo{person}{Jon
  Outland}, \bibinfo{person}{David King}, \bibinfo{person}{Jae~Kyu Lee},
  \bibinfo{person}{Ting-Peng Liang}, {and} \bibinfo{person}{Deborrah~C
  Turban}.} \bibinfo{year}{2017}\natexlab{}.
\newblock \bibinfo{booktitle}{\emph{Electronic commerce 2018: a managerial and
  social networks perspective}}.
\newblock \bibinfo{publisher}{Springer}.
\newblock
\urldef\tempurl%
\url{https://doi.org/10.1007/978-3-319-58715-8}
\showDOI{\tempurl}


\bibitem[\protect\citeauthoryear{WebAIM}{WebAIM}{2018}]%
        {webaim}
\bibfield{author}{\bibinfo{person}{WebAIM}.} \bibinfo{year}{2018}\natexlab{}.
\newblock \bibinfo{title}{Assistive Technologies}.
\newblock
\newblock
\newblock
\shownote{Retrieved July 12, 2018 from
  \url{https://webaim.org/articles/motor/assistive}.}


\bibitem[\protect\citeauthoryear{Wobbrock}{Wobbrock}{2014}]%
        {wobbrock2014improving}
\bibfield{author}{\bibinfo{person}{Jacob~O Wobbrock}.}
  \bibinfo{year}{2014}\natexlab{}.
\newblock \showarticletitle{Improving pointing in graphical user interfaces for
  people with motor impairments through ability-based design}.
\newblock In \bibinfo{booktitle}{\emph{Assistive Technologies and Computer
  Access for Motor Disabilities}}. \bibinfo{publisher}{IGI Global},
  \bibinfo{pages}{206--253}.
\newblock
\urldef\tempurl%
\url{https://doi.org/10.4018/978-1-4666-4438-0.ch008}
\showDOI{\tempurl}


\bibitem[\protect\citeauthoryear{Yim, Chu, and Sauer}{Yim
  et~al\mbox{.}}{2017}]%
        {yim2017augmented}
\bibfield{author}{\bibinfo{person}{Mark Yi-Cheon Yim},
  \bibinfo{person}{Shu-Chuan Chu}, {and} \bibinfo{person}{Paul~L Sauer}.}
  \bibinfo{year}{2017}\natexlab{}.
\newblock \showarticletitle{Is augmented reality technology an effective tool
  for e-commerce? An interactivity and vividness perspective}.
\newblock \bibinfo{journal}{\emph{Journal of Interactive Marketing}}
  \bibinfo{volume}{39} (\bibinfo{year}{2017}), \bibinfo{pages}{89--103}.
\newblock
\urldef\tempurl%
\url{https://doi.org/10.1016/j.intmar.2017.04.001}
\showDOI{\tempurl}


\bibitem[\protect\citeauthoryear{Zhang, Kulkarni, and Morris}{Zhang
  et~al\mbox{.}}{2017}]%
        {zhang2017smartphone}
\bibfield{author}{\bibinfo{person}{Xiaoyi Zhang}, \bibinfo{person}{Harish
  Kulkarni}, {and} \bibinfo{person}{Meredith~Ringel Morris}.}
  \bibinfo{year}{2017}\natexlab{}.
\newblock \showarticletitle{Smartphone-Based Gaze Gesture Communication for
  People with Motor Disabilities}. In \bibinfo{booktitle}{\emph{Proceedings of
  the 2017 CHI Conference on Human Factors in Computing Systems}}. ACM,
  \bibinfo{pages}{2878--2889}.
\newblock
\urldef\tempurl%
\url{https://doi.org/10.1145/3025453.3025790}
\showDOI{\tempurl}


\bibitem[\protect\citeauthoryear{Zhang and MacKenzie}{Zhang and
  MacKenzie}{2007}]%
        {zhang2007evaluating}
\bibfield{author}{\bibinfo{person}{Xuan Zhang} {and} \bibinfo{person}{I~Scott
  MacKenzie}.} \bibinfo{year}{2007}\natexlab{}.
\newblock \showarticletitle{Evaluating eye tracking with ISO 9241-part 9}. In
  \bibinfo{booktitle}{\emph{International Conference on Human-Computer
  Intrction}}. Springer, \bibinfo{pages}{779--788}.
\newblock
\urldef\tempurl%
\url{https://doi.org/10.1007/978-3-540-73110-8_85}
\showDOI{\tempurl}


\bibitem[\protect\citeauthoryear{Zuniga and Magee}{Zuniga and Magee}{2017}]%
        {zuniga2017camera}
\bibfield{author}{\bibinfo{person}{Rafael Zuniga} {and} \bibinfo{person}{John
  Magee}.} \bibinfo{year}{2017}\natexlab{}.
\newblock \showarticletitle{Camera Mouse: Dwell vs. Computer Vision-Based
  Intentional Click Activation}. In \bibinfo{booktitle}{\emph{International
  Conference on Universal Access in Human-Computer Interaction}}. Springer,
  \bibinfo{pages}{455--464}.
\newblock
\urldef\tempurl%
\url{https://doi.org/10.1007/978-3-319-58703-5_34}
\showDOI{\tempurl}


\end{thebibliography}

\end{document}